# Planetary and meteoritic Mg/Si and $\delta^{30}$Si variations inherited from solar nebula chemistry

Nicolas Dauphas[a,1], Franck Poitrasson[b], Christoph Burkhardt[a], Hiroshi Kobayashi[c], Kosuke Kurosawa[d]

[a]Origins Laboratory, Department of the Geophysical Sciences and Enrico Fermi Institute, The University of Chicago, 5734 South Ellis Avenue, Chicago, IL 60615, USA.

[b]Laboratoire Géosciences Environnement Toulouse, CNRS UMR 5563 – UPS – IRD, 14-16, avenue Edouard Belin, 31400 Toulouse, France.

[c]Department of Physics, Graduate School of Science, Nagoya University, Furo-cho, Chikusa-ku, Nagoya 464-8602, Japan

[d]Planetary Exploration Research Center, Chiba Institute of Technology, 2-17-1, Tsudanuma, Narashino, Chiba 275-0016, Japan

[1]To whom correspondence may be addressed. E-mail: dauphas@uchicago.edu

6688 words, 8 figures, 2 tables

Accepted in Earth and Planetary Science Letters

July 4, 2015

**Abstract**

**The bulk chemical compositions of planets are uncertain, even for major elements such as Mg and Si. This is due to the fact that the samples available for study all originate from relatively shallow depths. Comparison of the stable isotope compositions of planets and meteorites can help overcome this limitation. Specifically, the non-chondritic Si isotope composition of the Earth's mantle was interpreted to reflect the presence of Si in the core, which can also explain its low density relative to pure Fe-Ni alloy. However, we have found that angrite meteorites display a heavy Si isotope composition similar to the lunar and terrestrial mantles. Because core formation in the angrite parent-body (APB) occurred under oxidizing conditions at relatively low pressure and temperature, significant incorporation of Si in the core is ruled out as an explanation for this heavy Si isotope signature. Instead, we show that equilibrium isotopic fractionation between gaseous SiO and solid forsterite at ~1370 K in the solar nebula could have produced the observed Si isotope variations. Nebular fractionation of forsterite should be accompanied by correlated variations between the Si isotopic composition and Mg/Si ratio following a slope of ~1, which is observed in meteorites. Consideration of this nebular process leads to a revised Si concentration in the Earth's core of 3.6 (+6.0/-3.6) wt% and provides estimates of Mg/Si ratios of bulk planetary bodies.**

## 1. Introduction

The Earth's core displays a density deficit of 5-10 % relative to pure iron-nickel alloy. The three main contenders to explain this deficit are O, Si and S, with concentration estimates of 0-8 wt% for O, 2.8-12.5 wt% for Si, and 0-13 wt% for S (Hirose et al., 2013). Silicon is particularly appealing because its partitioning in the core would also explain why the Earth's mantle has a higher Mg/Si ratio than all documented chondrites (Wänke, 1981; Allègre et al., 1995). More recently, Si isotopes were proposed as a new tracer of the Si content of Earth's core (Georg et al., 2007). Measurements of terrestrial rocks and chondrites showed that the $\delta^{30}$Si value (deviation in permil of the $^{30}Si/^{28}Si$ ratio relative to the NBS-28 reference standard) of the bulk silicate Earth (BSE) was shifted by +0.15 ‰ or more relative to chondrites (Armytage et al., 2011; Fitoussi et al., 2009b; Georg et al., 2007). Experimental determinations and *ab initio* calculations of Si equilibrium isotopic fractionation between Si in silicate and metal showed that such a shift could be explained if a significant amount of Si was present in Earth's core (Georg et al., 2007; Hin et al., 2014; Shahar et al., 2011). Assuming that the bulk Earth has the same composition as ordinary and carbonaceous chondrites, the high Mg/Si ratio and $\delta^{30}$Si value of the BSE can be explained with ~8 to 12 wt% Si in the core.

A difficulty with this approach is that the isotopic compositions of several elements that display isotopic anomalies at a bulk planetary scale, such as O, Ca, Ti, Cr, Ni, and Mo, rule out derivation of the Earth from carbonaceous or ordinary chondrites and imply instead material more akin to enstatite chondrites (Dauphas et al., 2014a). Enstatite chondrites have very low Mg/Si ratios and low $\delta^{30}$Si values, which would call for an unrealistic large amount of Si in Earth's core (above ~20 wt%; Fitoussi and Bourdon, 2012; Zambardi et al., 2013) or the presence of an isotopically light (Huang et al., 2014) and Si-rich hidden reservoir in the deep Earth (Javoy et al., 2010).

We are therefore left with the difficulty that to estimate how much Si is in Earth's core from geochemical data (either Mg/Si ratios or $\delta^{30}$Si values), one has to know what the Earth is made of but none of the chondritic meteorite groups measured thus far can be taken as representative of the building blocks of the Earth. This motivated us to measure the Si isotope compositions of meteorite groups not studied before, in order to better understand what controls the Si isotope compositions and Mg/Si ratios of the bulk Earth and other planetary bodies.

Angrites are a group of basaltic achondrites that are among the most volatile-element depleted meteoritic samples available, with ratios of volatile-to-refractory elements Rb/Sr and K/U that are factors of ~150 and ~20 lower, respectively, than the terrestrial ratios (Davis, 2006). These samples are therefore well suited to evaluate the extent to which nebular and disk processes could have fractionated Si isotopes. In this contribution, we report the Si isotopic compositions of four angrites, along with ungrouped paired achondrites NWA 5363/NWA 5400 that share the O isotopic composition of the Earth-Moon system (Garvie, 2012; Weisberg et al., 2009; Burkhardt et al., 2015). Pringle et al. (2014) also reported Si isotope data of angrites and interpreted their measurements in terms of impact-induced Si volatilization on the angrite parent-body (APB). Based on our results (also see Dauphas et al., 2014b), we provide an alternative explanation and show that Mg/Si and $\delta^{30}$Si variations in angrites, chondrites and other planetary materials are best explained by equilibrium Si isotope fractionation between gaseous SiO and solid forsterite during condensation from nebular gas. This implies that Si isotopes are ambiguous tracers of Si in planetary cores but can be used to estimate Mg/Si ratios of bulk planetary bodies, a parameter that has a major influence on the mineral composition of rocky planets.

## 2. Samples and Methods

Four angrites samples with both quenched (D'Orbigny and Sahara 99555) and plutonic (NWA1670 and NWA6291) textures were analyzed. These samples are all finds and have experienced various degrees of terrestrial weathering. D'Orbigny is the least weathered, followed by Sahara 99555 and the two deserts "finds" NWA1670 and NWA6291. Angrites are differentiated meteorites of roughly basaltic composition that are composed mainly of Al, Ti-rich clinopyroxene (fassaite), Ca-bearing olivine, and anorthite (Mittlefehldt et al., 2002; Keil, 2012). These meteorites are highly depleted in the most volatile elements (*e.g.*, K and Rb) and are relatively enriched in more refractory elements like Ca and Ti. Angrites commonly serve as anchors in early solar system chronology and their ages were measured using various techniques, showing that quenched angrites crystallized ~4 Myr after solar system formation while plutonic angrites crystallized ~10 Myr after solar system formation (Kleine et al., 2012; Nyquist et al., 2009; Tang and Dauphas 2012, 2015). As their names indicate, quenched angrites have textures indicative of rapid cooling, with cooling rates ranging between 7 and 50 K/hr (Mikouchi et al., 2001). The iron isotopic composition of angrites is systematically shifted (~+0.1 ‰ in $\delta^{56}$Fe) relative to other meteorite groups (Wang et al., 2012). The reason for this shift is not well understood but could be related to their volatile-depleted nature.

NWA5363 and NWA5400 are paired ungrouped achondrites with a mineralogy and chemistry similar to brachinites (Garvie, 2012; Weisberg et al., 2009). NWA5400 contains ~79 % olivine ($Fa_{30}$), 10.5 % orthopyroxene, 8.9 % clinpyroxene, 1.4 % chromite, and other minor phases (Cl-rich apatite, troilite, kamacite, taenite) (Weisberg et al., 2009). NWA 5363 and NWA 5400 are interpreted to be residues of partial melting of a precursor similar in mineralogy and chemical composition to the R-chondrites (Gardner-Vandy et al., 2013). Both meteorites are affected by desert weathering, with NWA 5400 being more strongly altered than NWA 5363. Besides sampling the least weathered areas of both meteorites, a highly weathered part of NWA 5400 was also analyzed (NWA5400alt) to evaluate the effect of desert weathering on Si isotope compositions. The grading of weathering was confirmed by examination of the color of the powdered samples, which progresses to more reddish (more iron-oxidation) color in the sequence NWA 5363, 5400, and 5400alt. One motivation for analyzing these brachinite-like achondrites was that early work showed that they have $\Delta^{17}$O and $\varepsilon^{54}$Cr isotopic compositions very close to Earth (Weisberg et al., 2009; Shukolyukov et al., 2010; Garvie, 2012) and, unlike enstatite meteorites, they formed under relative high oxygen fugacity of IW-1 (Gardner-Vandy et al., 2013), thus raising the possibility that they might have been related to the building blocks of the Earth (Dauphas et al., 2014c). Subsequent work confirmed the Earth-like $\Delta^{17}$O but found that they had distinct isotopic compositions from the Earth for $\varepsilon^{54}$Cr and other elements, most notably $\varepsilon^{50}$Ti, so a simple relationship with Earth's building blocks is unlikely (Burkhardt et al., 2015).

Two chondrite “falls” (Allegan, H5; Pillistfer EL6) and a geostandard (BHVO-2) analyzed previously at the Géosciences Environnement Toulouse (GET) laboratory and other laboratories were also analyzed to compare the results with published datasets and ensure that the measurements were accurate.

Columns filled with Biorad AG50-X12 cation-exchange resin were used to purify Si from the samples after fusion with NaOH and dissolution (Georg et al., 2006). The typical Si recovery was at least 95 %. The procedural blank for Si was ~28 ng, which is less than 0.2 % of the Si signal and is thus negligible. A Thermo Electron Neptune MC-ICP-MS was used for isotope measurements in medium-resolution mode, under wet plasma condition, with the sample nebulized in a 0.05 M HCl solution following the approach described in Zambardi and Poitrasson (2011). Instrumental mass bias drift was corrected using both the sample standard bracketing technique and magnesium as an internal standard. Magnesium was added in both standards and samples, and the $^{25}Mg/^{24}Mg$ ratio was measured along with Si isotopes in the dynamic mode. Mass bias correction was applied using the exponential law (Russell et al., 1978). All data are reported as permil (‰) deviation of the $^{29}Si/^{28}Si$ ($\delta^{29}Si$) and $^{30}Si/^{28}Si$ ($\delta^{30}Si$) ratios of the samples relative to the NBS-28 international standard (Table 1). Nine analyses of BHVO-2 over the course of this study yielded $\delta^{30}Si$ = -0.262±0.053‰ (2SE), in good agreement within uncertainties with previous recent measurements (-0.286±0.011‰; average and 95% confidence interval of 14 measurements from Fitoussi et al., 2009; Savage et al., 2010, 2011; Zambardi and Poitrasson, 2011; Armytage et al., 2011; Armytage et al., 2012; Pringle et al., 2013; Savage and Moynier, 2013; Zambardi et al., 2013) Allegan and Pillistfer give $\delta^{30}Si$ values of -0.455±0.068‰ and -0.410±0.081‰, respectively. The $\delta^{30}Si$ value of Allegan was measured by Armytage et al. (2012) at -0.48±0.03‰, which is in good agreement with the result reported here. The $\delta^{30}Si$ value of Pillistfer was measured by Fitoussi and Bourdon (2012) at -0.58±0.03‰, which is lower than the result reported here. Savage and Moynier (2013) showed that the silicon isotopic composition of the various constituents of enstatite chondrites were variable, so the discrepancy between Fitoussi and Bourdon (2012) and this study may be due to non-representative sampling.

### 3. Results

Our new data reported in Table 1 and plotted in Fig. 1 fall on a mass-dependent fractionation line ($\delta^{30}Si \simeq 2\times\delta^{29}Si$). For this reason, we only discuss $\delta^{30}Si$ values in the following. NWA 5363, 5400, and 5400alt have identical $\delta^{30}Si$ values of -0.440±0.063, -0.502±0.050, and -0.443±0.060 ‰, indicating that desert weathering of these meteorites has little influence on bulk $\delta^{30}Si$ values. The weighted mean $\delta^{30}Si$ value of these three measurements is -0.468±0.033 ‰. The parent-body of NWA 5363/5400 may have had similar mineralogy and chemistry to the brachinite parent-body and R-chondrites (Gardner-Vandy et al., 2013) but there are no $\delta^{30}Si$ values in the literature for these meteorite groups. The $\delta^{30}Si$ of NWA 5363-5400 is close to the values obtained for carbonaceous and ordinary chondrites (Fig. 1), and is lower than the $\delta^{30}Si$ value of the BSE. While NWA 5363 and 5400 have identical $\Delta^{17}O$ and $\varepsilon^{62,64}Ni$ values to the Earth, they have different $\varepsilon^{48}Ca$, $\varepsilon^{50}Ti$, $\varepsilon^{54}Cr$, $\varepsilon^{92}Mo$ and $\varepsilon^{100}Ru$ values and thus cannot be the sole constituents of the Earth (Burkhardt et al., 2015).

The four analyzed angrites have the same Si isotopic composition within uncertainties (Fig. 1). The weighted mean $\delta^{30}$Si value of these four samples is -0.208±0.033 ‰. Unlike chondrites, which all have $\delta^{30}$Si values lower than the BSE, the $\delta^{30}$Si value of angrites appears to be slightly higher than the BSE value. Pringle et al. (2014) measured the Si isotope composition of five angrites and found significant variations among these samples, with $\delta^{30}$Si values that range between -0.42±0.04 and -0.23±0.04 ‰. The most negative value in their dataset overlaps with the values of ordinary chondrites and carbonaceous chondrites, while the highest values overlap with the BSE composition. D'Orbigny is the only angrite analyzed by both groups and its $\delta^{30}$Si value of -0.240±0.068 ‰ reported in Table 1 is heavier than the value of -0.36±0.04 ‰ reported by Pringle et al. (2014). Pringle et al. (2014) report a mean $\delta^{30}$Si value for angrites of -0.33±0.12 ‰. This value falls between the value of ordinary/carbonaceous chondrites of ~-0.45 and the BSE value of ~-0.29 ‰, and it overlaps within error with both. Instead, our results show that angrites have a $\delta^{30}$Si value that is very distinct from those of ordinary and carbonaceous chondrites.

## 4. Discussion

Existing results on meteorites and planetary bodies provide some context for our data. Bulk chondrites display variations in their Mg/Si ratios (Mg/Si ratios are expressed as atomic ratios throughout this paper) (Wasson and Kallemeyn, 1988) and $\delta^{30}$Si values (Fitoussi et al., 2009; Armytage et al., 2011; Fitoussi and Bourdon, 2012; Pringle et al., 2013; Savage and Moynier, 2013; Zambardi et al., 2013) (Fig. 2). Enstatite chondrites have low Mg/Si ratios (EH=0.73; EL=0.88) and low $\delta^{30}$Si values (EH=-0.690±0.051 ‰; EL=-0.581±0.033 ‰). Savage and Moynier (2013) found that the non-magnetic portions of enstatite chondrites were isotopically heavier than the bulk. The origin of this intrinsic heterogeneity is uncertain and could have involved nebular as well as parent-body processes such as equilibrium isotopic fractionation between coexisting phases during metamorphism. Aubrites, which sample the silicate portion of a differentiated enstatite-chondrite-like parent-body also have low $\delta^{30}$Si values (-0.582±0.053 ‰). Removal of Si into the core of the aubrite parent-body would have shifted the $\delta^{30}$Si value of the silicate portion to higher values, so the $\delta^{30}$Si value of aubrites must be an upper-limit on the bulk value of their parent-body. This is further evidence that $\delta^{30}$Si values were variable at the scale bulk planetary bodies and that the low $\delta^{30}$Si and Mg/Si values documented in EH and EL are representative of large-scale nebular reservoirs. Compared to enstatite chondrites and aubrites, ordinary chondrites have higher Mg/Si ratios (H=0.96; L=0.93; LL=0.94) and higher $\delta^{30}$Si values (H=-0.460±0.032 ‰; L=-0.467±0.028 ‰; LL=-0.430±0.061 ‰). Carbonaceous chondrites have still higher Mg/Si ratios (CI=1.07; CM=1.05; CO=1.05; CV=1.07; CK=1.10; CH=1.06) but $\delta^{30}$Si values that are close to ordinary chondrites (CI=-0.440±0.112 ‰; CM=-0.485±0.090 ‰; CO=-0.458±0.054 ‰; CV=-0.435±0.041 ‰; CK=-0.410±0.112 ‰; CH=-0.450±0.046 ‰). The $\delta^{30}$Si values for Mars and Vesta as defined by SNC and HED meteorites are -0.48±0.03 ‰ and -0.42±0.03 ‰, respectively, and are in the range of values measured in ordinary and carbonaceous chondrites (Fig. 1). Overall, chondrites define a trend in $\delta^{30}$Si *vs*. Mg/Si that is heavily controlled by EH and EL chondrites (Fitoussi et al., 2009; Dauphas et al., 2014a) (Fig. 2).

Terrestrial igneous rocks show variations in $\delta^{30}$Si values associated with igneous differentiation that are most noticeable in samples that contain more than ~49 wt% $SiO_2$ (Savage et al., 2011, 2014; Poitrasson and Zambardi, 2015). The average $\delta^{30}$Si of magmatic igneous rocks with less than 49 wt% $SiO_2$ (n=20; 49 wt% is a conservative cutoff based on a visual examination of $\delta^{30}$Si *vs*. $SiO_2$ values of igneous rocks) and mantle peridotites (n=33) yields a mean $\delta^{30}$Si value for the silicate Earth of -0.297±0.025 (Fitoussi et al., 2009; Armytage et al., 2011; Savage et al., 2010, 2011; Zambardi and Poitrasson, 2011; Savage and Moynier, 2013; Zambardi et al., 2013). Lunar rocks have indistinguishable $\delta^{30}$Si value compared to the Earth (-0.292±0.026 ‰; n=38) (Armytage et al., 2012; Fitoussi and Bourdon, 2012; Zambardi et al., 2013). Both the BSE and the Moon have heavy Si isotopic compositions relative to all chondrite groups measured thus far. The BSE also has a high Mg/Si ratio of ~1.25 (Wänke, 1981; Allègre et al., 1995; McDonough and Sun, 1995). In a $\delta^{30}$Si *vs*. Mg/Si plot, the BSE composition lies on the correlation defined by bulk chondrites (Fig. 2).

The heavy Si isotopic composition of the silicate Earth relative to chondrites was interpreted to reflect equilibrium isotopic fractionation between Si in metal and silicate during core formation, opening the possibility of constraining how much Si is in the core (Georg et al., 2007; Fitoussi et al., 2009; Ziegler et al., 2010; Armytage et al., 2011; Shahar et al., 2011; Zambardi et al., 2013; Hin et al., 2014; Savage et al., 2014). However, a major difficulty with this interpretation is that the silicon isotopic composition of the material that made the Earth is unknown. Assuming that the bulk Earth has an enstatite chondrite composition leads to estimates of the amount of Si in Earth's core that are unrealistically high, above 20 wt% (Fitoussi and Bourdon, 2012; Zambardi et al., 2013). Carbonaceous and ordinary chondrites are ruled out as main constituents of the Earth because they display isotopic anomalies relative to terrestrial rocks for O, Ti, Ca, Cr, Ni, and Mo (Dauphas et al., 2014a,c). The finding that angrites have $\delta^{30}$Si values similar to or higher than the BSE (Fig. 1) calls into question previous interpretations that heavy Si isotopic signatures in planetary mantles are signatures of core formation. Below, we propose an explanation as to why angrites have a heavy Si isotopic composition, and discuss the implications of nebular fractionation on the Mg/Si ratio and the composition of planets.

4.1. Magmatic differentiation?

Terrestrial rocks with high silica content can have their silicon isotopic compositions modified by magmatic differentiation (Savage et al., 2011, 2014). The $\delta^{30}$Si of terrestrial magmatic rocks correlates roughly with the $SiO_2$ concentration; $\delta^{30}Si=0.0056[wt\%SiO_2]-0.567$ (Savage et al., 2014). However, basalts have $\delta^{30}$Si values indistinguishable from mantle peridotites, suggesting that basalts are representative of their mantle source (Fitoussi et al., 2009; Savage et al., 2010; Savage et al., 2011; Armytage et al., 2011; Zambardi and Poitrasson, 2011; Savage and Moynier, 2013; Zambardi et al., 2013). Further evidence that the Si isotopic compositions of mafic igneous rocks are minimally affected by magmatic processes comes from SNC and HED meteorites (Fitoussi et al., 2009; Armytage et al., 2011; Pringle et al., 2013; Zambardi et al., 2013), which comprise rocks of basaltic compositions that have Si isotopic compositions very similar to carbonaceous and ordinary chondrites, but very distinct from angrites. Hence, based on available evidence, partial melting and magmatic differentiation is an unlikely cause for the heavy Si isotope composition of angrites and their $\delta^{30}$Si value may be taken as representative of the whole APB mantle.

4.2. Silicon partitioning in the core of the angrite parent-body?

For Si to become siderophile, partitioning must happen under either high temperature (*i.e.*, great depth where high P-T conditions are encountered) or low oxygen fugacity conditions. Equilibrium silicon isotope fractionation between metal and silicate during planetary core formation has been the dominant framework through which the heavy Si isotopic composition of the Earth's mantle was explained (Georg et al., 2007; Fitoussi et al., 2009; Ziegler et al., 2010; Armytage et al., 2011; Shahar et al., 2011; Zambardi et al., 2013; Hin et al., 2014; Savage et al., 2014). The +0.15 ‰ heavier Si isotopic composition of the BSE relative to ordinary and carbonaceous chondrites was interpreted to reflect the presence of ~8 to 12 wt% Si in Earth's core. Pringle et al. (2013) also interpreted a potential shift between Vesta and carbonaceous chondrites of ~+0.1 ‰ as evidence that at least 1% of Si partitioned into the core of Vesta, corresponding to a $fO_2$ of IW-4. However, under such low $fO_2$, little $Fe^{2+}$ would be left in Vesta's mantle. Indeed, assuming ideality, the FeO content of the mantle is related to the oxygen fugacity through $X_{\mathrm{FeO}}/X_{\mathrm{Fe}} = 10^{\Delta\mathrm{IW}/2} = 0.01$ and FeO/(MgO+FeO)≃0. This is inconsistent with the composition of HED meteorites and remote observations of Vesta, which give $X_{\mathrm{FeO}}/X_{\mathrm{Fe}} = 0.24$ and FeO/(MgO+FeO)=0.32 (Toplis et al., 2014), suggesting that other processes are at play.

Likewise, silicon core partitioning is not a viable explanation for the heavy Si isotope composition of angrites because core formation in the APB did neither occur at high temperature nor low oxygen fugacity. Quenched angrites are very well dated by several radiometric dating techniques that give an age of around ~4 Myr after the start of the solar system formation (Kleine et al., 2012; Nyquist et al., 2009; Tang and Dauphas, 2012, 2015). For such pristine igneous rocks to survive from the earliest epochs of the solar system, the body that they came from must have been small, as otherwise protracted magmatism and large impacts would have wiped out this early magmatic history, as is the case for the Moon and Earth. Scott and Bottke (2011) reviewed evidence available to constrain the size of the APB and they concluded that it was most likely between 100 and 500 km in size, which for an average density of 3 300 kg/m$^3$ corresponds to a central pressure of only 0.04-1 kbar. In planetesimals of such size, heating by $^{26}$Al-decay produces temperatures that are not expected to exceed ~1400 °C (*i.e.*, below the liquidus; Neumann et al., 2014). Righter (2008) and Shirai et al. (2009) estimated the $fO_2$ during core formation on the angrite parent-body to have been around IW-1 based on concentration measurements of redox-sensitive, moderately siderophile elements such as Ni, Co, Mo and W.

At a pressure of 0.04-1 kbar, low temperature of ~1400 °C, and high $fO_2$ of IW-1, silicon would not partition in any significant amount in the core of the angrite parent-body (Ziegler et al., 2010). This is a robust conclusion, so core-partitioning of Si can be ruled out as an explanation of the heavy Si isotopic composition of angrites relative to chondrites.

4.3. Impact volatilization?

Pringle et al. (2014) measured $\delta^{30}Si$ values in angrites that ranged between the compositions of ordinary/carbonaceous chondrites and the BSE. These authors explained the overall heavier Si isotope composition of angrites relative to carbonaceous/ordinary chondrites by impact-induced volatilization. Large impacts such as the Moon forming-impact release considerable energy, leading to large-scale melting and partial volatilization (Tonks and Melosh, 1993; Canup, 2004). Such volatilization could possibly explain the depletion of the Moon in volatile elements such as potassium. Large impacts were also invoked to explain the Fe and Zn stable isotopic fractionations in the Earth-Moon system (Poitrasson et al., 2004; Paniello et al., 2012). As discussed below, on smaller objects, the consequences of impacts would be less significant.

Collisions between planetary bodies induce shock compression with a local increase in pressure and entropy. After impact, the pressure in the shocked material is released but no heat is lost to the surrounding medium because the impact timescale is short compared to the timescale of heat transport. Assuming furthermore that the process is reversible, the entropy during decompression can be considered constant (adiabatic+reversible=isentropic). This provides a means of estimating the amount of vapor produced in an impact by comparing the calculated entropy of the shocked material with the low-pressure (~1 bar) entropies of incipient vaporization ($S_{\mathrm{iv}}$) and complete vaporization ($S_{\mathrm{cv}}$). The vapor fraction ($\psi$) is given by the lever rule (Ahrens and O'Keefe, 1972; Kurosawa et al., 2012),

$$\begin{aligned} &\psi = 0 \ \text{ for } \ S_0 + \delta S < S_{\mathrm{iv}}, \\ &\psi = (S_0 + \delta S - S_{\mathrm{iv}})/(S_{\mathrm{cv}} - S_{\mathrm{iv}}) \ \text{ for } \ S_{\mathrm{iv}} \le S_0 + \delta S \le S_{\mathrm{cv}}, \qquad (1) \\ &\psi = 1 \ \text{ for } \ S_0 + \delta S > S_{\mathrm{cv}}, \end{aligned}$$

where $S_0$ and $\delta S$ are the starting entropy and the entropy increase by shock compression, respectively. The main difficulty in this approach is to calculate the entropy increase $\delta S$ in the shocked material, as this requires an accurate knowledge of the equation of state over a wide range of P-T conditions. Sugita et al. (2012) presented a simplified approach to calculate the entropy along the Hugoniot (locus of final shocked states) using the following parameters: initial density, bulk sound velocity, relationship between particle and shockwave velocities, and Grüneisen parameters (Appendix A). Integration of $\psi$ over the volume of shocked material (Eq. A6) gives the mass fraction of vapor in a given impact, $f_{\mathrm{vap}}$, as a function of the impact velocity $v_{\mathrm{imp}}$,

$$f_{\mathrm{vap}} = M_{\mathrm{vap}}\left(v_{\mathrm{imp}} \sin\theta\right)/M_{\mathrm{imp}}, \qquad (2)$$

where $\theta$ is the impact angle and $M_{\mathrm{imp}}$ is the mass of the impactor (projectile). The $\sin\theta$ dependence comes from the fact that to first order, impact melting and vaporization is controlled by the normal component of the impact velocity (Perazzo and Melosh, 2000). Ahrens and O'Keefe (1972) quantified the effect of impacts on rock volatilization and concluded that the fraction vaporized was small unless impact velocities exceeded ~10 km/s. However, this threshold is only relevant to materials that are solid. Much of the material accreted in the first ~3 Myr of the formation of the solar system would have acquired enough $^{26}Al$ to melt from the heat produced by decay of this short-lived radionuclide (Grimm and McSween, 1993; Dauphas and Chaussidon, 2011). It is thus conceivable that impacts involved molten rather than solid bodies. In Fig. 3A, the vaporized fraction is plotted as a function of impact velocity for collisions involving molten bodies and the threshold for vaporization is reduced to ~6 km/s. To compute the net effect of vaporization for different final body sizes, the vaporized fractions were folded in a statistical model of planetary accretion that tracks the masses and velocities of the bodies involved in collisional accretion (Kobayashi et al., 2010; Kobayashi and Dauphas, 2013).

During runaway and oligarchic growth, collisional coagulation of planetesimals produced Moon to Mars-sized embryos. The collisional velocity is given by $\sqrt{v_{esc}^2 + v_r^2}$, where $v_{esc}$ is the escape velocity and $v_r$ is the random velocity $v_r^2 \simeq (e^2 + i^2)v_K^2$, with $e$, $i$, and $v_K$ the eccentricity, inclination, and keplerian velocity, respectively. Because orbital eccentricities and inclinations were suppressed by gas drag (Kobayashi et al., 2010), impact velocities between bodies of radii $r_1$, $r_2$ and masses $m_1$, $m_2$ were set by the mutual escape velocities, $v_{imp} \approx v_{esc} = \sqrt{2\mathcal{G}\,(m_1 + m_2)/(r_1 + r_2)}$. The fraction of material that experienced impact-induced volatilization relative to the total mass of a body is given by,

$$F = (1/M) \int_0^M f_{\mathrm{vap}}(v_{\mathrm{esc}}(m) \sin\theta)\, dm, \quad (3)$$

where $v_{esc}(m) = \sqrt{2\mathcal{G}\, m/r} = (2\mathcal{G})^{1/2}(4\pi\rho/3)^{1/6}m^{1/3}$ because the growing body of mass $m$ mainly accretes much smaller bodies ($\rho$ is the density). As shown in Fig. 3B, $F$ is negligible for $m < 0.2\, M_{\mathrm{Earth}}$ because for such size objects, the escape velocity is lower than the 6 km/s threshold. The parent-body of angrites would have been too small ($v_{esc} = 0.25$ km/s for a 500 km size body) to experience significant impact volatilization. To summarize, theoretical considerations do not support impact as a cause for Si isotope fractionation in planetesimals.

4.4. Equilibrium isotopic fractionation during forsterite condensation

Meteorites show variable Mg/Si ratios that may reflect chemical fractionation in the solar nebula/protoplanetary disk. Theory predicts that forsterite condensation should have been a major control on Mg/Si fractionation (Larimer and Anders, 1970; Yoneda and Grossman, 1995). Forsterite is the only phase in the condensation sequence that was abundant enough and had sufficiently fractionated Mg/Si ratio (2×solar) to change the Mg/Si ratio at a bulk planetary scale. As discussed below, an inevitable consequence of forsterite fractionation is the fractionation of silicon isotopes. We will show that equilibrium isotope fractionation between forsterite and gaseous SiO can explain the $\delta^{30}$Si *vs*. Mg/Si correlation found in different chondrite parent-bodies and the Earth, together with their constant $\delta^{26}$Mg values (Fig. 2).

At high nebular temperatures, silicon and magnesium are in the gas phase as $Mg_g$ and $SiO_g$ ($SiS_g$ under reducing conditions) with a solar (CI) Mg/Si ratio of ~1.07 (Yoneda and Grossman, 1995). Upon cooling, forsterite (Mg/Si≃2) condenses and thereby lowers the Mg/Si ratio of the remaining gas. At ~1370 K (for a total pressure of $10^{-3}$ bar), when forsterite has finished condensing, nearly all Mg is in forsterite, while half of the Si is still in the gas (Fig. 4). Let us now examine the consequences of these simple considerations for Mg and Si isotopes. As a first approach, the gas is assumed to remain in isotopic equilibrium with the solid until forsterite is fully condensed. Clayton et al. (1978) first calculated the extent of equilibrium Si stable isotope fractionation between forsterite and $SiO_{gas}$ and concluded that this fractionation should be about 2 ‰ at the temperatures relevant to forsterite condensation, a value that has been confirmed by more recent studies. Based on spectroscopic data and first principle calculations, the equilibrium fractionation factors between forsterite-$Mg_{gas}$ (Huang et al., 2013; Schauble, 2011) and

forsterite-$SiO_{gas}$ (Javoy et al., 2012; Méheut et al., 2009; Pahlevan et al., 2011) are (with $T$ in K; Fig. 5),

$$\Delta^{26}\mathrm{Mg}_{\text{Forsterite-Mg(gas)}} = 2.2\times10^6/T^2, \quad (4)$$

$$\Delta^{30}\mathrm{Si}_{\text{Forsterite-SiO(gas)}} = 4.2\times10^6/T^2. \quad (5)$$

At 1370 K, the temperature corresponding to the end of forsterite condensation, equilibrium fractionations between solid and gas would be +1.2 ‰ for $\delta^{26}$Mg and +2.3 ‰ for $\delta^{30}$Si, respectively. Note that the fractionation factor should be similar for forsterite-$SiS_g$ under reducing condition (Javoy et al., 2012). The isotopic mass-balance between gas and solid for element X (Mg or Si) can be written as,

$$\delta\mathrm{X}_{\text{gas}} = \delta\mathrm{X}_{\text{CI}} - \left(1 - f_{\text{Xgas}}\right)\Delta\mathrm{X}_{\text{solid-gas}}, \quad (6)$$

$$\delta\mathrm{X}_{\text{solid}} = \delta\mathrm{X}_{\text{CI}} + f_{\text{Xgas}}\Delta\mathrm{X}_{\text{solid-gas}}, \quad (7)$$

where $f_{\text{Xgas}}$ is the fraction of $X$ in the gas and $\delta\mathrm{X}_{\text{gas}}$, $\delta\mathrm{X}_{\text{solid}}$, $\delta\mathrm{X}_{\text{CI}}$ are the isotopic compositions of $X$ in the gas, solid, and CI chondrites (taken as proxy for solar composition: $\delta^{26}\mathrm{Mg}_{\mathrm{CI}}=-0.27$ ‰, Teng et al., 2010; $\delta^{30}\mathrm{Si}_{\mathrm{CI}}=-0.44$ ‰, Armytage et al., 2011; Fitoussi et al., 2009b; Zambardi et al., 2013).

When forsterite is fully condensed, much of the initial Mg is in forsterite $f_{\text{Mg(gas)}} = 0$, so the solid inherits the isotopic composition of the starting gas while the small amount of gas remaining is shifted by the gas-solid equilibrium isotopic fractionation: $\delta^{26}\mathrm{Mg}_{\text{forsterite}}=-0.27$ ‰ and $\delta^{26}\mathrm{Mg}_{\text{gas}}=-0.27-1.2=-1.47$ ‰. At the same time, approximately half of Si is in the gas and the rest is in forsterite $f_{\text{Si(gas)}} = 0.5$, so the equilibrium isotopic fractionation between solid and gas is expressed about equally in the two phases: $\delta^{30}\mathrm{Si}_{\text{forsterite}}=-0.44+2.3\times0.5=+0.71$ ‰ and $\delta^{30}\mathrm{Si}_{\text{gas}}=-0.44-2.3\times0.5=-1.59$ ‰.

The forsterite and $SiO_g$ reservoirs could have then remixed in different proportions, separated by dust-gas decoupling or planet formation, and interacted by reaction of $SiO_g$ with forsterite to make enstatite. The curves expected for mixing between the gas and forsterite end-members for $\delta^{26}$Mg-Mg/Si and $\delta^{30}$Si-Mg/Si are:

$$\delta^{30}\mathrm{Si}_{\text{mixture}} = x\delta^{30}\mathrm{Si}_{\text{gas}} + (1-x)\delta^{30}\mathrm{Si}_{\text{solid}} \quad (8)$$

$$\delta^{26}\mathrm{Mg}_{\text{mixture}} = \frac{x(\mathrm{Mg/Si})_{\text{gas}}\delta^{26}\mathrm{Mg}_{\text{gas}} + (1-x)(\mathrm{Mg/Si})_{\text{solid}}\delta^{26}\mathrm{Mg}_{\text{solid}}}{x(\mathrm{Mg/Si})_{\text{gas}} + (1-x)(\mathrm{Mg/Si})_{\text{solid}}} \quad (9)$$

where $x$ is the fraction of Si from the gas, which depends on the Mg/Si ratio of the mixture through,

$$x = \frac{(\mathrm{Mg/Si})_{\text{solid}} - (\mathrm{Mg/Si})_{\text{mixture}}}{(\mathrm{Mg/Si})_{\text{solid}} - (\mathrm{Mg/Si})_{\text{gas}}} \quad (10)$$

The more complex formula for $\delta^{26}$Mg compared to $\delta^{30}$Si arises from the fact that mixing in a $\delta^{26}$Mg-Mg/Si diagram has a curvature that depends on the Mg/Si ratios of the mixture end-members (Langmuir et al., 1978).

In Fig. 6a,b, we plot mixing curves between the forsterite and gas end-members calculated above. For $\delta^{30}$Si *vs*. Mg/Si, the mixing curve is a straight line of slope ~1.1, which reproduces well the observed correlation given the assumptions made (slope=0.85±0.16). The lack of correlation between $\delta^{26}$Mg and Mg/Si (Bourdon et al., 2010; Teng et al., 2010) is also well explained because the mixing curve between forsterite and gas has a strong curvature and little variation is expected for the $\delta^{26}$Mg values of meteorites. The reason for this is that although the gas has a very low $\delta^{26}$Mg, by the time forsterite has fully condensed, little Mg is left in the gas, so the $\delta^{26}$Mg value of gas-solid mixtures is buffered by forsterite.

The simple model presented above explains the correlated variations between $\delta^{30}$Si and Mg/Si. A number of simplifications and assumptions were made that we scrutinize hereafter. One such simplification is that forsterite is not the only Mg-bearing phase that condenses from a gas of solar composition. Thermodynamic calculations show that other Mg-Si bearing phases also condense at higher and lower temperatures (*e.g.*, mellilite, diopside, anorthite, enstatite, spinel). To take this into account, we have calculated the compositions of the solid and gas end-members using the model output of Yoneda and Grossman (1995). The predicted mixing curves are similar to those calculated above using the assumption that forsterite is the only condensing silicate (Fig. 6c,d). However, one should bear in mind that the more refined thermodynamic calculation may still miss some important aspects of the condensation sequence. For instance, below ~1370 K, gaseous SiO starts reacting with forsterite to make enstatite. Because enstatite has a Mg/Si ratio of ~1, this decreases the Mg/Si ratio of the bulk condensed solid and steepens the $\delta^{30}$Si-Mg/Si and $\delta^{26}$Mg-Mg/Si correlations. However, this reaction may be kinetically inhibited because a layer of enstatite would form at the surface of forsterite grains that would prevent the reaction from proceeding, thereby extending the stability range of forsterite (Imae et al., 1993). In this respect, the simple model presented above may capture adequately what happened in the solar nebula.

The calculations above were made under the assumption of equilibrium isotopic fractionation. Experiments and theory have shown that during evaporation and condensation, significant kinetic isotopic fractionation may be present (Davis et al., 1990; Young et al., 1998; Wang et al., 1999; Richter et al., 2002, 2007, 2009; Richter, 2004; Dauphas et al., 2004; Dauphas and Rouxel, 2006; Shahar and Young, 2007). During condensation, the light isotopes impinge surfaces at a higher rate than heavier isotopes, so condensed phases tend to be enriched in the light isotopes. This process is best described using the Hertz-Knudsen equation, from which one can show that the degree of equilibrium and kinetic isotope fractionation depends on the degree of supersaturation of the surrounding medium (Appendix B, Eq. B24),

$$\Delta_{\text{Condensation}} = \frac{P_{\text{sat}}}{P}\Delta_{\text{Equilibrium}} + \left(1 - \frac{P_{\text{sat}}}{P}\right)\Delta_{\text{Kinetic}}, \quad (11)$$

where $P_{\text{sat}}$ is the equilibrium saturation vapor pressure of the element considered, $P$ is its vapor pressure, $\Delta_{\text{Equilibrium}}$ is the equilibrium isotopic fractionation (in ‰) between solid and gas, and $\Delta_{\text{Kinetic}} = 1000\left(\gamma_i/\gamma_j\sqrt{m_j/m_i} - 1\right)$ with $\gamma$ the condensation coefficients and $m$ the mass of the isotope chemical species $i$ and $j$ in the gas. At 1370 K, the equilibrium fractionation factors were calculated above (Eqs. 4 and 5) and we have $\Delta^{26}\text{Mg}_{\text{Equilibrium}}^{\text{Fo/Mg-gas}} = +1.2‰$ and $\Delta^{30}\text{Si}_{\text{Equilibrium}}^{\text{Fo/SiO-gas}} = +2.3‰$. For the kinetic isotope fractionation factor, we assume that the condensation coefficients are identical for the different isotopologues and use the masses of the gas species (Mg for magnesium, SiO for silicon). It follows that $\Delta^{26}\text{Mg}_{\text{Kinetic}}^{\text{Fo/Mg-gas}} = 1000\left(\sqrt{24/26} - 1\right) = -39‰$ and $\Delta^{30}\text{Si}_{\text{Kinetic}}^{\text{Fo/Mg-gas}} = 1000\left(\sqrt{44/46} - 1\right) = -22‰$. In Fig. 7, we plot the isotopic fractionation associated with condensation when both equilibrium and kinetic isotope effects are taken into account. Either $P_{\mathrm{sat}}/P = 1$ and there is significant equilibrium isotope fractionation between forsterite and gaseous Mg or SiO, or $P_{\mathrm{sat}}/P < 1$ and there is significant kinetic isotope fractionation associated with condensation. This calculation shows that isotopic fractionation of Si during forsterite condensation is inevitable, unless all Si was condensed, which would not explain the variations in Mg/Si of bulk chondrites. Kinetic effects during forsterite condensation would reduce the slope of the correlation between $\delta^{30}$Si and Mg/Si (Fig. 6) because condensation in a kinetic regime is expected to enrich the condensate in the light isotopes, which is opposite to the effect expected for equilibrium condensation.

Fitoussi et al. (2009) explained the low $\delta^{30}$Si value of enstatite chondrite by assuming the presence of kinetic isotope fractionation associated with the chemical reaction of gaseous SiO and forsterite to make enstatite. However, the authors ruled out this process as an explanation for the heavy Si isotopic composition of the BSE (and by extension the angrite parent-body) because it could only decrease the Mg/Si and $\delta^{30}$Si values from solar (CI chondrite) composition. However, the starting composition would not be CI but rather the forsterite composition calculated above, which is shifted in its $\delta^{30}$Si value by ~+1 ‰ relative to CI and has a high Mg/Si ratio ~2×CI. We have shown above that the gas left after equilibrium condensation of forsterite has low $\delta^{30}$Si and physical mixing alone would have been sufficient to explain the $\delta^{30}$Si-Mg/Si correlation (Fig. 6). Nevertheless, because of kinetic inhibition of enstatite condensation, significant silicon supersaturation could have developed when the system entered the stability field of enstatite, which could have fractionated Si isotopes. The reacting SiO would have lower $\delta^{30}$Si than the equilibrium value calculated above and if anything, this kinetic effect would produce a steeper $\delta^{30}$Si -Mg/Si slope.

The lack of large stable isotope variations for elements more volatile than silicon was taken as an argument against volatility-controlled silicon isotope fractionation in the Earth (Fitoussi et al., 2009; Hin et al., 2014). However, the comparison may be misleading because volatile element fractionation would occur at much lower temperatures (*e.g.*, ~1000 K for potassium; 700 K for Zn) than Si (~1370 K), at different times and places. Furthermore, we are not aware of previous detailed studies of gas-solid equilibrium isotope fractionation factors for volatile elements, so it is unknown if any variation would be detectable given current analytical uncertainties.

Forsterite fractionation is thought to be responsible for Mg/Si variations in undifferentiated meteorites (Larimer and Anders, 1970) and the present work suggests that it is also responsible for fractionation of Si isotopes. A question arises as to why forsterite condensation was so significant in shaping the composition of planets. Forsterite represents a large fraction of the condensable matter and is solely responsible for changing the solar nebula from a low-dust to a high-dust environment between 1440 and 1370 K (Fig. 4). Such a large change in the dust content must have undoubtedly promoted dust aggregation. Furthermore, a thermostatic effect must have accompanied forsterite condensation (Morfill, 1988). Energy from viscous dissipation is deposited at the midplane and is transported to the disk photosphere where it is radiated away. Local cooling of the disk would have led to forsterite condensation and increase in the dust opacity, thereby hampering vertical heat loss (and vice-versa). While the temperature window of forsterite condensation is rather narrow (1440-1370 K), the thermostatic effect provided ample time for forsterite to be decoupled from the gas by aerodynamic processes, such as gas drag, turbulent concentration, and settling onto the mid-plane.

4.5. Implications for the compositions of planetary bodies and their cores

The calculations presented above demonstrate that fractionation of forsterite in the nebula can explain the range of Mg/Si and $\delta^{30}$Si values measured in meteorites and planets. In this respect, the high Mg/Si and $\delta^{30}$Si values of the BSE may be representative of the bulk Earth, simply reflecting the greater incorporation of high-temperature forsterite in its constituents. Those same constituents may have formed the Moon-forming impactor, explaining why lunar and terrestrial rocks have identical $\delta^{30}$Si values (Dauphas et al., 2014a), which was difficult to explain in the context of silicon isotopic fractionation during core formation as in most giant impact scenarios, most of the Moon is thought to have come from the impactor (Canup, 2004; however see Ćuk and Stewart 2012; Canup 2012) and Earth-Moon equilibration scenarios predicted that the $\delta^{30}$Si value of the Moon should be shifted relative to the BSE by ~0.4 ‰ (Pahlevan et al., 2011)._

If our interpretation is correct, then one can use the $\delta^{30}$Si values of the silicate portions of differentiated bodies to calculate the Mg/Si ratio of their bulk parent-bodies, as mantle melting fractionates Mg/Si ratios but does not seem to affect $\delta^{30}$Si values (Savage et al., 2011, 2014). Except for the Earth's mantle, the major element chemistry of planetary objects are very poorly known because our sampling is limited to shallow depths and the rocks available for sampling are not representative of the underlying mantles. An example of that is given by the Moon, for which estimates of atomic Mg/Si ratios vary from 0.89 to 1.27 (Dauphas et al., 2014a). In Table 2 and Fig. 8, we report Mg/Si ratios calculated using the regression of chondrite data from Fig. 2, which is controlled by the low Mg/Si and $\delta^{30}$Si values of enstatite chondrites. For all bodies except the Earth and the aubrite parent-body, we assume that all silicon is in the mantle. For Earth, we calculate the intersection between the nebular trend defined by chondrites and the calculated bulk Earth compositions defined by the BSE and experimental metal-silicate fractionation for silicon isotopes (Hin et al., 2014) (Fig. 8). The silicon concentration in Earth's core estimated by this approach is 3.6 (+6.0/-3.6) wt% and is consistent with core composition models that call for the presence of elements other than silicon to explain the density deficit of Earth's core (Hirose et al., 2013; Badro et al., 2014). Our results and interpretations show that an important virtue of Si isotopes that has been hitherto unexplored is its potential to constrain bulk Mg/Si ratios of differentiated objects, which has important bearings on the constitution of planets, since it influences the nature of the mineral phases that can be found at depth, such as the proportions of olivine and pyroxene.

## 5. Conclusion

New isotope measurements of angrites reveal a $\delta^{30}$Si for the angrite parent-body that is similar or heavier than the bulk silicate Earth value. The only plausible explanation for the heavy Si isotope composition of angrites is nebular fractionation, most likely associated with forsterite condensation. In the condensation sequence, forsterite is the only mineral that is abundant enough and possesses a high enough Mg/Si ratio to induce significant variation in the Mg/Si ratio of bulk chondrites, as is observed. The equilibrium $\delta^{30}$Si fractionation between forsterite and nebular SiOg is ~2 ‰ at the temperatures relevant to forsterite condensation. Removal or addition of such forsterite condensate should be associated with a correlation between $\delta^{30}$Si and Mg/Si, which is observed in chondrites.

Variations in Mg/Si and $\delta^{30}$Si of chondrites define a correlation that goes through the BSE composition. The finding of high $\delta^{30}$Si values in angrites demonstrates that nebular fractionation can produce bodies with Mg/Si and $\delta^{30}$Si values like the BSE, with no need to invoke partitioning of large amounts of Si in Earth's core. This new interpretation leads to a downward revision of the amount of Si in Earth's core to 3.6 (+6.0/-3.6) wt%, which is consistent with the presence of at least another element than Si in Earth's core.

Silicon isotopes provide limited constraints on the amount of Si in planetary cores. Instead, they are best used as proxies of the Mg/Si ratios of bulk planetary planets, a parameter that largely influences the mineralogical composition, melting regime, and rheology of planetary mantles.

**ACKNOWLEDGEMENTS**. J. Chmeleff, M. Henry and J. Prunier maintain the MC-ICP-MS from Observatoire Midi-Pyrénées and Géoscience Environnement Toulouse clean lab facilities in excellent working order. L. Grossman, A.J. Campbell, F.J. Ciesla, R. Fischer, P.H. Warren, Philipp Heck, and T. Zambardi are thanked for their advice and discussions. The meteorite specimens were purchased from meteorite dealers or were generously provided by the Robert A. Pritzker Center for Meteoritics (Field Museum) and the Smithsonian Institution. FP was supported by CNRS. ND was supported by grants from NSF (EAR 1144429, 1444951, 1502591) and NASA (NNX12AH60G, NNX14AK09G, NNX15AJ25G).

## Appendix A.

The pressure away from the isobaric core is given by,

$$P(r) = P_0 \left(\frac{r}{r_{ic}}\right)^{-\alpha}, \qquad \text{(A1)}$$

where $r$ is the distance from the impact point, $r_{ic}$ is the size of isobaric core, which is assumed to be equal to the radius of the projectile, $P_0$ is the pressure in the isobaric core and $\alpha$ is a constant. We set $\alpha = 3$ because this choice is reasonable for a three-dimensional shock propagation (Melosh, 1989). For the projectile, $P = P_0$ is used. The shock velocity, $U_s$, is related to the particle velocity ($U_p$; velocity of the shocked material in the initial reference frame) by a linear relationship,

$$U_s = C_0 + sU_p, \qquad \text{(A2)}$$

where $C_0$ and $s$ are the bulk sound velocity and a constant, respectively. From Eq. (A2) and the Rankine-Hugoniot relations it follows,

$$P(R) = U_p(R)\left[C_0 + sU_p(R)\right]\rho_0, \qquad \text{(A3)}$$

where $\rho_0$ is the initial bulk density.

Using Eq. (A2), the differential form of Rankine-Hugoniot relations and the Grüneisen equation of state, Sugita et al. (2012) showed that the following differential equations apply,

$$\frac{dS}{dU_p} = \frac{sU_p^2}{T(C_0+sU_p)}, \qquad \text{(A4)}$$

$$\frac{dT}{dU_p} = \frac{C_0\Gamma_0 T\left(C_0+sU_p-U_p\right)^{q-1}}{\left(C_0+sU_p\right)^{q+1}} + \frac{\frac{1}{C_v}sU_p^2}{C_0+sU_p}, \qquad \text{(A5)}$$

where $S$, $T$, $\Gamma_0$, and $C_{\rm v}$ are the entropy, temperature, Grüneisen parameter, and constant-volume heat capacity, respectively. Integration of Eqs. (A4) and (A5) from $U_p = 0$ to $U_p$ gives the entropy gain $\delta S(r)$.

The vapor mass is calculated from $\delta S(r)$ as,

$$M_{\rm vap} = \psi\left(S_0 + \delta S(r_{ic})\right)M_{\rm projectile} + \int \psi\left(S_0 + \delta S(r)\right) m_{in}(r)\mathrm{d}r, \qquad \text{(A6)}$$

where $m_{in}(r)\mathrm{d}r$ is the mass in the range from $r$ to $r+\mathrm{d}r$, $\psi$ is determined as $\psi = 0$ for $S_0 + \delta S < S_{\rm iv}$, $\psi = (S_0 + \delta S - S_{\rm iv})/(S_{\rm cv} - S_{\rm iv})$ for $S_{\rm iv} \le S_0 + \delta S \le S_{\rm cv}$, and $\psi = 1$ for $S_0 + \delta S > S_{\rm cv}$. The parameters relevant to cold basalt and molten basalt are shown in the table below.

| | "Cold" basalt at 293 K | Molten basalt at 1393 K |
|---|---|---|
| Starting density, $\rho_{\rm o}$ [Mg/m$^3$] | 2.86[a] | 2.61[f] |
| Bulk sound velocity, $C_{\rm o}$ [km/s] | 3.5[b] | 3.06 ($u_{\rm p}$ < 1.5 km/s)[f] |
| | | 0.85 ($u_{\rm p}$ > 1.5 km/s)[f] |
| $s = dU_s/dU_p$ | 1.3[b] | 1.36 ($u_{\rm p}$ < 1.5 km/s)[f] |
| | | 2.63 ($u_{\rm p}$ > 1.5 km/s)[f] |
| Grüneisen parameter, $\Gamma_{\rm o}$ | 1.39[c] | 1.52[g] |
| Power-law exponent, $q$ | 1[d] | 1.6[g] |
| Isochoric specific heat, $C_{\rm v}$ | 1.15[e] | 1.522[e] |
| [kJ/K/kg] | | |
| Starting entropy, $S_o$ [kJ/K/kg] | 0.527 [c] | 2.682 [c] |
| Incipient vaporization entropy, $S_{iv}$ [kJ/K/kg] | 3.461 [c] | 3.461 [c] |
| Complete vaporization entropy, $S_{cv}$ [kJ/K/kg] | 7.654 [c] | 7.654 [c] |

a. Melosh (1989). b. Sekine et al. (2008). c. Ahrens and O'Keefe (1972). d. Assumed. e. Bouhifd et al. (2007). f. Rigden et al. (1988). g. Stixrude et al. (2009).

## Appendix B.

The Hertz-Knudsen equation gives the net condensation or evaporation flux $J_i$ (in moles per unit area per unit time) of a chemical species $i$ as a function of the difference between the partial ($P_i$) and saturation ($P_{i,\text{sat}}$) vapor pressures (a next flux from the condensed phase to the gas is counted positively),

$$J_i = \frac{n_i \gamma_i (P_{i,\text{sat}} - P_i)}{\sqrt{2\pi m_i RT}}, \quad \text{(B1)}$$

where $n_i$ is the number of atoms of $i$ in the gas species molecule, $\gamma_i$ is the evaporation/condensation coefficient, $m_i$ is the molar mass of the gas species containing $i$, $R$ is the gas constant, and $T$ is the temperature in K. If one considers two isotopes $i$ and $j$, the ratio of the fluxes is (Richter, 2004; Richter et al., 2002),

$$\frac{J_i}{J_j} = \alpha_{\text{Kin}} \frac{P_{i,\text{sat}} - P_i}{P_{j,\text{sat}} - P_j}, \quad \text{(B2)}$$

where $\alpha_{\text{Kin}}$ is the kinetic fractionation factor,

$$\alpha_{\text{Kin}} = \frac{\gamma_i}{\gamma_j}\sqrt{\frac{m_j}{m_i}}. \quad \text{(B3)}$$

Using the Hertz-Knudsen equation, Richter et al. (2002, 2007) and Richter (2004) derived the equations that give the isotopic fractionation for (i) evaporation for any value of $P/P_{\text{sat}}$ when there is no equilibrium fractionation (Eq. 13 of Richter et al., 2002), (ii) evaporation when $P/P_{\text{sat}} = 0$ and there is equilibrium fractionation (Eq. 6 of Richter et al., 2007), and (iii) condensation for any value of $P/P_{\text{sat}}$ when there is no equilibrium fractionation (Eq. 31 of Richter 2004). Both Young et al. (1998) and Wang et al. (1999) considered the case of diffusion-limited evaporation from a single component system. Richter et al. (2002; 2011) derived the equations for the more relevant case of diffusion-limited evaporation of the more volatile components from a multi-component melt. Diffusion-limited transport both in the condensed phase and surrounding gas is neglected below. We follow the approach of Dauphas and Rouxel (2006) to derive expressions that give the isotope fractionations during evaporation and condensation for any value of $P/P_{\text{sat}}$ when there is significant equilibrium isotope fractionation.

a. Evaporation

We consider that the condensed phase is well homogenized and define the fractionation factor as fractionation between the condensed phase and evaporated atoms ($N_{\text{cp}}$ is the number of atoms in the condensed phase),

$$\alpha_{\text{Evap.}}=\frac{N_{i,\text{cp}}/N_{j,\text{cp}}}{J_i/J_j}.\ \text{(B4)}$$

Eq. B2 can be rewritten as,

$$\frac{J_i}{J_j}=\alpha_{\text{Kin}}\frac{P_{i,\text{sat}}}{P_{j,\text{sat}}}\left(\frac{1-P_i/P_{i,\text{sat}}}{1-P_j/P_{j,\text{sat}}}\right).\qquad \text{(B5)}$$

The equilibrium fractionation factor between condensed phase and gas (assumed to be ideal) is given by,

$$\alpha_{\text{Eq}}=\frac{N_{i,\text{cp}}/N_{j,\text{cp}}}{P_{i,\text{sat}}/P_{j,\text{sat}}}.\qquad \text{(B6)}$$

Dividing the two sides of Eq. B5 by $N_{i,\text{cp}}/N_{j,\text{cp}}$ it follows,

$$\frac{1}{\alpha_{\text{Evap.}}}=\frac{\alpha_{\text{Kin}}}{\alpha_{\text{Eq}}}\left(\frac{1-P_i/P_{i,\text{sat}}}{1-P_j/P_{j,\text{sat}}}\right),\qquad \text{(B7)}$$

which can be rewritten as,

$$\frac{1}{\alpha_{\text{Evap.}}}=\frac{\alpha_{\text{Kin}}}{\alpha_{\text{Eq}}}\left(\frac{1-\frac{P_i/P_j}{P_{i,\text{sat}}/P_{j,\text{sat}}}\times P_j/P_{j,\text{sat}}}{1-P_j/P_{j,\text{sat}}}\right).\qquad \text{(B8)}$$

We assume that the isotopic composition of the gas phase surrounding the sample is set by the material being evaporated (see Richter et al., 2002, 2011, for a discussion of the effect of diffusive transport in the gas),

$$P_i/P_j=J_i/J_j.\qquad \text{(B9)}$$

Combining Eqs. B4, B6, and B9, we have,

$$\frac{P_i/P_j}{P_{i,\text{sat}}/P_{j,\text{sat}}}=\frac{J_i/J_j}{P_{i,\text{sat}}/P_{j,\text{sat}}}=\alpha_{\text{Eq}}\frac{J_i/J_j}{N_{i,\text{cp}}/N_{j,\text{cp}}}=\frac{\alpha_{\text{Eq}}}{\alpha_{\text{Evap.}}}\qquad \text{(B10)}$$

Eq. B8 can therefore be rewritten as,

$$\frac{1}{\alpha_{\text{Evap.}}}=\frac{\alpha_{\text{Kin}}}{\alpha_{\text{Eq}}}\left(\frac{1-\frac{\alpha_{\text{Eq}}}{\alpha_{\text{Evap.}}}P_j/P_{j,\text{sat}}}{1-P_j/P_{j,\text{sat}}}\right).\ \text{(B11)}$$

This equation can be solved for $\alpha_{\text{Evap.}}$,

$$\alpha_{\text{Evaporation}}=\frac{1+\frac{P_j}{P_{\text{j,sat}}}(\alpha_{\text{Kin}}-1)}{\alpha_{\text{Kin}}/\alpha_{\text{Eq}}}\simeq\frac{1+\frac{P}{P_{\text{sat}}}(\alpha_{\text{Kin}}-1)}{\alpha_{\text{Kin}}/\alpha_{\text{Eq}}}.\qquad \text{(B12)}$$

Taking the logarithm of this expression and recognizing that $ln(1+x)\simeq x$ when $x$ is small, we have in $\Delta=1000(\alpha-1)\simeq 1000ln\alpha$ notation,

$$\Delta_{\text{Evaporation}}=\Delta_{\text{Equilibrium}}-\left(1-\frac{P}{P_{\text{sat}}}\right)\Delta_{\text{Kinetic}}.\qquad \text{(B13)}$$

with

$$\Delta_{\text{Evaporation}}=\delta_{\text{Condensed phase}}-\delta_{\text{Evaporated atoms}},\qquad \text{(B14)}$$

$$\Delta_{\text{Kinetic}}=1000\left(\frac{\gamma_i}{\gamma_j}\sqrt{\frac{m_j}{m_i}}-1\right),\qquad \text{(B15)}$$

$$\Delta_{\text{Equilibrium}}=\left[\delta_{\text{Condensed phase}}-\delta_{\text{gas}}\right]_{\text{Equilibrium}}.\qquad \text{(B16)}$$

a. Condensation

During condensation, the fractionation factor between the condensing atoms and the gas phase is defined as ($N_{\text{gas}}$ is the number of atoms in the gas phase),

$$\alpha_{\text{Cond.}}=\frac{J_i/J_j}{N_{i,\text{gas}}/N_{j,\text{gas}}}=\frac{J_i/J_j}{P_i/P_j}.\qquad \text{(B17)}$$

Eq. B2 can be rewritten as,

$$\frac{J_i}{J_j} = \alpha_{\mathrm{Kin}} \frac{P_i}{P_j} \left( \frac{P_{i,\mathrm{sat}}/P_i - 1}{P_{j,\mathrm{sat}}/P_j - 1} \right), \qquad \text{(B18)}$$

$$\alpha_{\mathrm{Cond.}} = \alpha_{\mathrm{Kin}} \left( \frac{P_{i,\mathrm{sat}}/P_i - 1}{P_{j,\mathrm{sat}}/P_j - 1} \right). \qquad \text{(B19)}$$

The saturation vapor pressure is defined relative to the composition of the atoms at the surface of the condensed phase, so the equilibrium fractionation factor between the condensed atoms and gas is given by,

$$\alpha_{\mathrm{Eq}} = \frac{J_i/J_j}{P_{i,\mathrm{sat}}/P_{j,\mathrm{sat}}}. \qquad \text{(B20)}$$

Eq. B19 can be rewritten as,

$$\alpha_{\mathrm{Cond.}} = \alpha_{\mathrm{Kin}} \left( \frac{\frac{P_{i,\mathrm{sat}}/P_{j,\mathrm{sat}}}{P_i/P_j} P_{j,\mathrm{sat}}/P_j - 1}{P_{j,\mathrm{sat}}/P_j - 1} \right). \qquad \text{(B21)}$$

Introducing Eq. B17 and B20 in this expression, it follows that,

$$\alpha_{\mathrm{Cond.}} = \alpha_{\mathrm{Kin}} \left( \frac{\frac{\alpha_{\mathrm{Cond.}}}{\alpha_{\mathrm{Eq}}} P_{j,\mathrm{sat}}/P_j - 1}{P_{j,\mathrm{sat}}/P_j - 1} \right). \qquad \text{(B22)}$$

This equation can be solved for $\alpha_{\mathrm{Cond.}}$,

$$\alpha_{\mathrm{Cond.}} = \frac{\alpha_{\mathrm{Kin}}}{1 + \frac{P_{j,\,\mathrm{sat}}}{P_j} \left( \frac{\alpha_{\mathrm{Kin}}}{\alpha_{\mathrm{Eq}}} - 1 \right)} \simeq \frac{\alpha_{\mathrm{Kin}}}{1 + \frac{P_{\mathrm{sat}}}{P} \left( \frac{\alpha_{\mathrm{Kin}}}{\alpha_{\mathrm{Eq}}} - 1 \right)}. \qquad \text{(B23)}$$

In $\Delta$ notation, this takes the form,

$$\Delta_{\mathrm{Condensation}} = \frac{P_{\mathrm{sat}}}{P} \Delta_{\mathrm{Equilibrium}} + \left(1 - \frac{P_{\mathrm{sat}}}{P}\right) \Delta_{\mathrm{Kinetic}}. \qquad \text{(B24)}$$

with

$$\Delta_{\mathrm{Condensation}} = \delta_{\mathrm{Condensed\ atoms}} - \delta_{\mathrm{Gas}}, \qquad \text{(B25)}$$

and $\Delta_{\mathrm{Kinetic}}$ and $\Delta_{\mathrm{Equilibrium}}$ defined as in Eqs. B15, B16.

## Figure Captions

**Figure 1.** Silicon ($\delta^{30}$Si relative to NBS 28) isotopic compositions of angrites, brachinite-like ungrouped achondrite NWA 5363/5400 (Table 1) and planetary reservoirs (see text for details and references). Angrites have $\delta^{30}$Si values similar to the BSE and Moon compositions and heavier than other planetary bodies. PB, EL, EH, OC, CC, HED, and BSE stand for parent-body, EL and EH-type enstatite chondrites, ordinary chondrites, carbonaceous chondrites, Howardite-Eucrite-Diogenite achondrites, and bulk silicate Earth, respectively.

**Figure 2**. Silicon ($\delta^{30}$Si relative to NBS 28) and magnesium ($\delta^{26}$Mg relative to DSM) isotope compositions *vs.* atomic Mg/Si ratios of chondrites (enstatite, ordinary, and carbonaceous, noted EC, OC, CC, respectively; green triangle=EH, green square=EL, orange square=L, orange triangle=LL, orange circle=H, black square=CM, black diamond=CO, black right-pointing triangle= CH, black circle= CI, black down-pointing triangle=CV, black up-pointing triangle=CK) and the bulk silicate Earth (BSE) with their 95% uncertainties. The Mg/Si ratios of the mantles of Mars (parent-body of the Shergotty-Nakhla-Chassigny meteorites – SNC), Moon, Vesta (parent-body of the Howardite-Encrite-Diogenite meteorites – HED) and the aubrite, angrite, ureilite and NWA5363 parent-bodies are uncertain, so their $\delta^{30}$Si values are plotted off the Mg/Si scale on the right. **a**. The grey band is the best-fit line to the chondrite data with 68% prediction interval; $\delta^{30}\text{Si}=0.85\times(\text{Mg/Si})_{\text{atomic}}-1.32$. The thick blue solid line is a model prediction (slope 1.15) for mixing between forsterite (Mg/Si=2; $\delta^{30}$Si=0.71) and nebular gas (Mg/Si=0; $\delta^{30}$Si=-1.59) at ~1370 K. This shows that nebular fractionation can account for the $\delta^{30}$Si and Mg/Si variations documented in planetary materials, including angrites and the Earth. The green curve shows possible bulk Earth trajectories based the composition of the BSE and on experimental calibration of Si isotopic fractionation between metal and silicate (Hin et al., 2014) for different values of Si in the core (0-12 wt.%). **b**. Similar diagram for magnesium. As predicted by the model, no variation in magnesium isotopic composition is observed among chondrites and the silicate Earth. Mg/Si of the BSE and chondrites are from Wasson and Kallemeyn, (1988), McDonough and Sun (1995). $\delta^{30}$Si values are from this study and Georg et al., (2007), Fitoussi et al. (2009), Savage et al. (2010); Armytage et al. (2011); Zambardi and Poitrasson (2011); Armytage et al. (2012); Fitoussi and Bourdon (2012); Savage and Moynier (2013); Pringle et al. (2013); Zambardi et al. (2013); $\delta^{26}$Mg values are from Teng et al. (2010).

**Figure 3.** Volatilization of planetary bodies by impacts. A. Fraction of vapor, $f_{\text{vap}}$ (Eq. 2), generated in a single collision between two dissimilar-size bodies with impact velocity $v_{\text{imp}}$, for "cold" basalt (blue) and molten basalt (red). B. Vaporization fraction, $F$ (Eq. 3), during the growth history of a body with mass $M$ for "cold" basalt (blue) and molten basalt (red) (the parameters of this calculation are given in Appendix A). The dotted curves represent $F$ for the runaway and oligarchic growth, setting $\theta$ as the most likely impact angle of 45° (see text for details). During later stages, growth occurs through giant impacts, which occur in small enough numbers that adopting a single impact angle is unwarranted. Down-pointing triangles ($\theta = 60°$), circles ($\theta = 45°$), and up-pointing triangles ($\theta = 30°$) indicate $F$ for the giant impact stage. The late stage of planetary growth was modeled as collisions with bodies of 0.1 $M_{Earth}$ and the maximum collisional velocity for merging from Kokubo and Genda (2010) was used. The parameters of this calculation are given in Appendix A.

**Figure 4**. Distribution of Mg and Si between nebular gas and condensed phases *vs*. temperature (in K) in a gas of solar composition at a total pressure of $10^{-3}$ atm. Reproduced after Fig. 1 in Davis and Richter (2014) (also see Yoneda and Grossman, 1995). At high nebular temperatures, silicon and magnesium are in the gas phase as $Mg_g$ and $SiO_g$ ($SiS_g$ under reducing conditions). During cooling, forsterite (Mg/Si=2) condenses and thereby lowers the Mg/Si of the remaining gas. At ~1370 K, when forsterite has finished condensing, nearly all Mg is in forsterite, while half of the Si is still in the gas. Fractionation and mixing of the condensed forsterite and the depleted nebular gas can explain the variable Mg/Si ratios and $\delta^{30}$Si values among planetary bodies as well as their constant $\delta^{26}$Mg values.

**Figure 5**. Equilibrium Mg and Si isotopic fractionations between forsterite, gaseous Mg and SiO (Eqs. 4, 5; calculated using the fractionation factors from Clayton et al., 1978; Méheut et al., 2009; Schauble, 2011; Pahlevan et al., 2011; Javoy et al., 2012; Huang et al., 2013). Gaseous Mg and SiO are the relevant gas species for condensation of solar gas at $10^{-3}$ atmosphere (Yoneda and Grossman, 1995). The vertical lines at 1440 and 1370 K correspond approximately to the beginning and end of forsterite condensation (Fig. 4).

**Figure 6**. Silicon ($\delta^{30}$Si relative to NBS 28) and magnesium ($\delta^{26}$Mg relative to DSM) isotope compositions *vs.* Mg/Si ratios of chondrites (EC, OC, CC) and the terrestrial mantle (BSE) with their 95 % confidence intervals. **a**. The solid line is a model prediction (slope 1.15) for mixing between forsterite (Mg/Si=2; $\delta^{30}$Si=0.71) and nebular gas (Mg/Si=0; $\delta^{30}$Si=-1.59) at ~1370 K (see main text for details). This shows that nebular fractionation can account for the $\delta^{30}$Si and Mg/Si variations documented in planetary materials, including angrites and the Earth. **b**. Similar diagram for magnesium. As predicted by the model, no variation in magnesium isotopic composition is observed among chondrites and the silicate Earth. **c,d.** Same as panels a and b except that the end-members are not assumed to be pure forsterite and Mg-free nebular gas but rather condensate and gas compositions at 1370 K from a model of condensation of solar gas at $10^{-3}$ bar total pressure (Yoneda and Grossman 1995; Fig. 4). See Fig. 2 caption and main text for references.

**Figure 7.** Calculated Mg and Si isotopic fractionations during condensation as a function of $P_{sat}/P$ (Appendix B, Eqs. 11, B24). When $P_{sat}/P=1$, the system is at equilibrium and the only isotopic fractionation expressed is the equilibrium one. When $P_{sat}/P<1$, some kinetic isotopic fractionation is present that is maximal when $P_{sat}/P=0$. Forsterite condensed from nebular gas is expected to be isotopically fractionated relative to the gas. For Mg, this is not manifested in bulk samples because Mg condenses quantitatively in forsterite. The kinetic fractionation balances the equilibrium fractionation for $\delta^{30}$Si or $\delta^{26}$Mg when $P_{\text{sat}}(\text{SiO})/P(\text{SiO}) = 0.91$ and $P_{\text{sat}}(\text{Mg})/P(\text{Mg}) = 0.97$, respectively. However, there is no reason to consider that these *ad hoc* conditions would have been met in the solar nebula.

**Figure 8**. $\delta^{30}$Si of planetary bodies *vs*. their inferred atomic bulk Mg/Si ratios (Table 2). Grey band and dark green curves represent the best-fit line to the chondrite data with its 68% prediction interval for the regression of Mg/Si on $\delta^{30}$Si, and trajectories for the bulk Earth composition for 0-12 wt.% Si in the core, respectively (*cf*. Fig. 2). Mg/Si ratios were calculated under the assumption of no Si in planetary cores, except for Earth and the aubrite parent-body. For Earth, the Mg/Si estimate is derived from the intersection of the chondrite regression and the bulk Earth metal/silicate fractionation trajectory using an experimental calibration (Hin et al., 2014). This yields a terrestrial Si core concentration of $3.6^{+6}_{-3.6}$ wt%. For the aubrite parent-body, some Si may be present in the core but this cannot be reliably quantified because the Mg/Si ratio of the mantle of the aubrite parent-body is not known. Uncertainties on inferred bulk planetary Mg/Si ratios are at the 68 % confidence level.

**Table 1.** Silicon isotope compositions of meteorites and a terrestrial reference basalt

| Sample | $\delta^{30}$Si (‰) | 95% c.i.[a] | $\delta^{29}$Si (‰) | 95% c.i.[a] | *n* |
|---|---|---|---|---|---|
| NWA 5363 | -0.440 | 0.063 | -0.243 | 0.026 | 6 |
| NWA 5400 | -0.502 | 0.050 | -0.244 | 0.019 | 6 |
| NWA 5400alt | -0.443 | 0.060 | -0.246 | 0.085 | 6 |
| *Angrites* | | | | | |
| d'Orbigny | -0.240 | 0.068 | -0.121 | 0.057 | 6 |
| NWA 1670 | -0.168 | 0.052 | -0.084 | 0.019 | 6 |
| Sahara 99555 | -0.243 | 0.068 | -0.096 | 0.045 | 3 |
| NWA 6291 | -0.221 | 0.076 | -0.111 | 0.061 | 6 |
| *Chondrites* | | | | | |
| Allegan H5 | -0.455 | 0.068 | -0.204 | 0.045 | 3 |
| Pillistfer EL6 | -0.410 | 0.081 | -0.243 | 0.051 | 6 |
| *Terrestrial standard* | | | | | |
| BHVO-2[b] | -0.262 | 0.053 | -0.132 | 0.037 | 9 |

$\delta^{29,30}\mathrm{Si}=\left[({}^{29,30}\mathrm{Si}/{}^{28}\mathrm{Si})_{\mathrm{Sample}}/({}^{29,30}\mathrm{Si}/{}^{28}\mathrm{Si})_{\mathrm{NBS\text{-}28}}-1\right]\times 10^{3}$ , where NBS-28 is a reference material. Details on the analytical methods are provided in Zambardi and Poitrasson (2011). [a]Errors are given as 95 % confidence intervals of the means of *n* analyses. Allegan and Sahara 99555 were measured 3 times, which is too small to derive reliable error bars from the dispersion of the data. For this reason, we use the reproducibility of the other samples measured during the same session as a measure of error. [b]Values in good agreement with previously published results (Savage et al., 2010; Armytage et al., 2011; Zambardi et al., 2013).

**Table 2.** Silicon isotope compositions and inferred atomic Mg/Si ratios of bulk planetary bodies

| Planetary body | $\delta^{30}$Si (‰)[a] | Mg/Si[b] | Si in core (wt%)[c] |
|---|---|---|---|
| Angrite parent body | -0.208 ± 0.033 | 1.27 ± 0.09 | 0 |
| Moon | -0.292 ± 0.013 | 1.21 ± 0.08 | 0 |
| Earth | -0.334 ± 0.012 | $1.16^{+0.10}_{-0.13}$ | $3.6^{+6.0}_{-3.6}$ |
| HED parent body (Vesta) | -0.420 ± 0.015 | 1.06 ± 0.07 | 0 |
| NWA5363 parent body | -0.468 ± 0.017 | 1.00 ± 0.06 | 0 |
| Ureilite parent body | -0.47 ± 0.05 | 1.00 ± 0.07 | 0 |
| Mars | -0.480 ± 0.017 | 0.99 ± 0.06 | 0 |
| Aubrite parent body | <-0.582 ±0.053 | <0.94 | ? |

All uncertainties are at the 68 % confidence level.

[a] $\delta^{30}$Si is measured in mantle-derived rocks (see text for references), except for Earth, where $\delta^{30}$Si is obtained from the intersection between the regression of chondrite data and the experimental metal/silicon equilibrium trajectory (Hin et al., 2014) originating from the BSE ($\delta^{30}$Si=-0.297±0.025 ‰; see text for references). For the aubrite parent-body, some Si may be in the core and the $\delta^{30}$Si value measured in aubrites represents an upper-limit on the $\delta^{30}$Si value of the bulk object, so we are only able to set an upper-limit on the Mg/Si ratio of the aubrite parent-body.

[b] Bulk Mg/Si ratio obtained by projecting $\delta^{30}$Si values of individual planetary bodies onto the chondrite data regression $\delta^{30}$Si=0.85×(Mg/Si)-1.32.

[c] A Si core content of 0 is assumed for all bodies other than Earth and the aubrite parent-body. The terrestrial value is obtained from the intersection of chondrite data regression and experimental metal/silicate equilibrium fractionation trajectory originating

from the BSE.

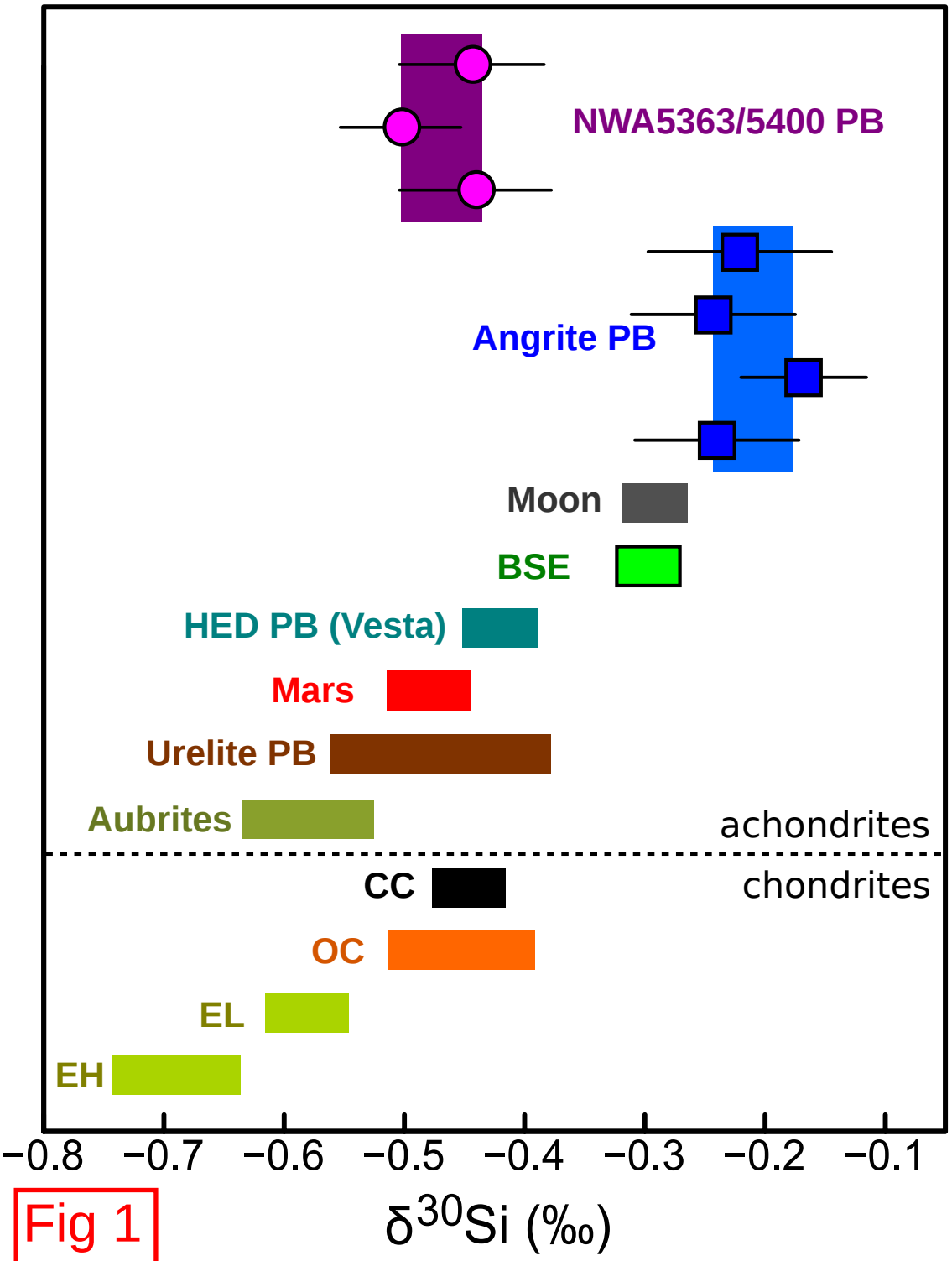
NWA5363/5400 PB
Angrite PB
Moon
BSE
HED PB (Vesta)
Mars
Urelite PB
Aubrites
achondrites
chondrites
CC
OC
EL
EH
−0.8
−0.7
−0.6
−0.5
−0.4
−0.3
−0.2
−0.1
δ30Si (‰)

Fig 1

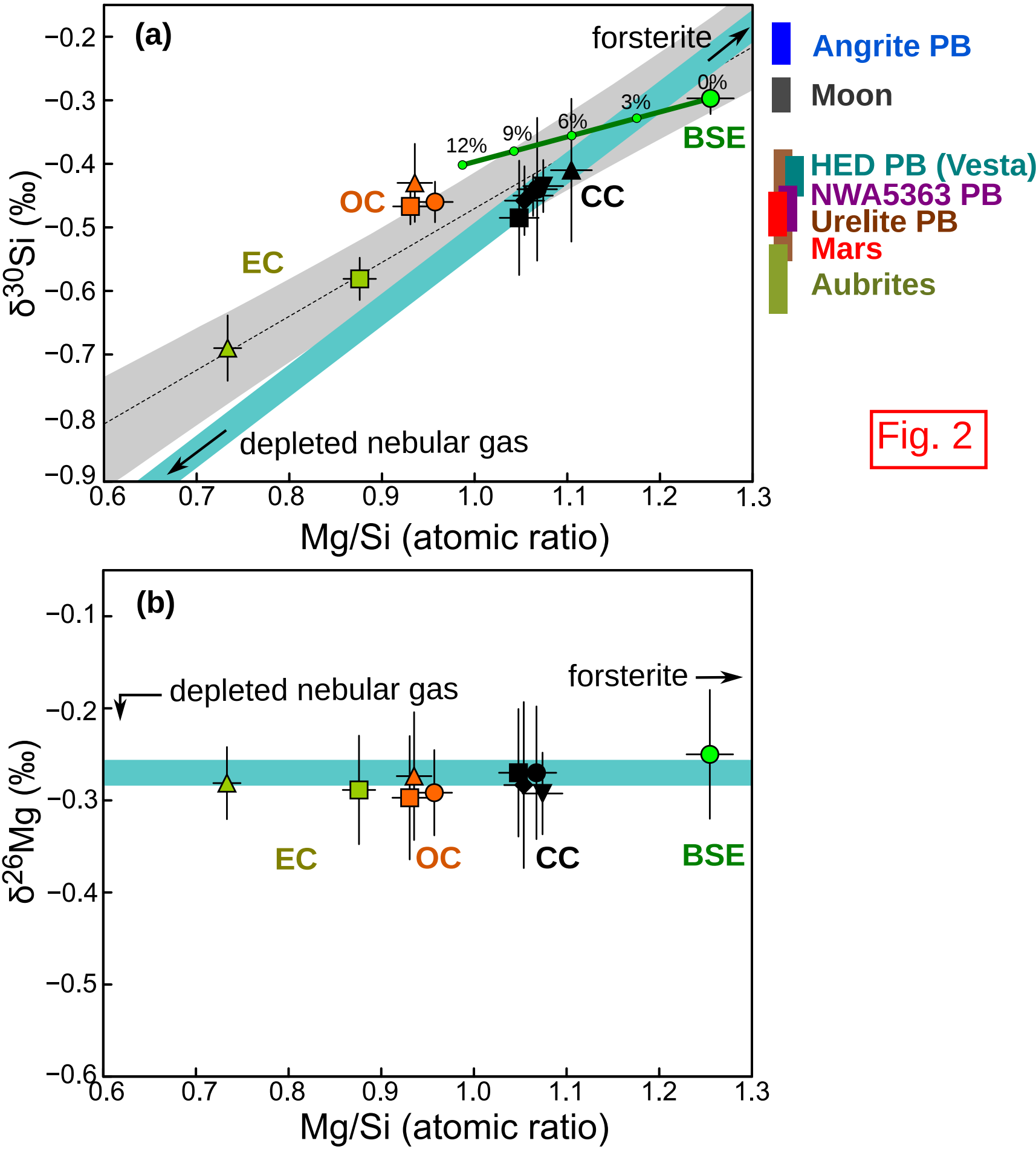

(a)
forsterite
Angrite PB
Moon
0%
3%
6%
9%
12%
BSE
HED PB (Vesta)
NWA5363 PB
Urelite PB
Mars
Aubrites
OC
CC
EC
depleted nebular gas
δ30Si (‰)
Mg/Si (atomic ratio)
(b)
depleted nebular gas
forsterite
EC
OC
CC
BSE
δ26Mg (‰)
Mg/Si (atomic ratio)

Fig. 2

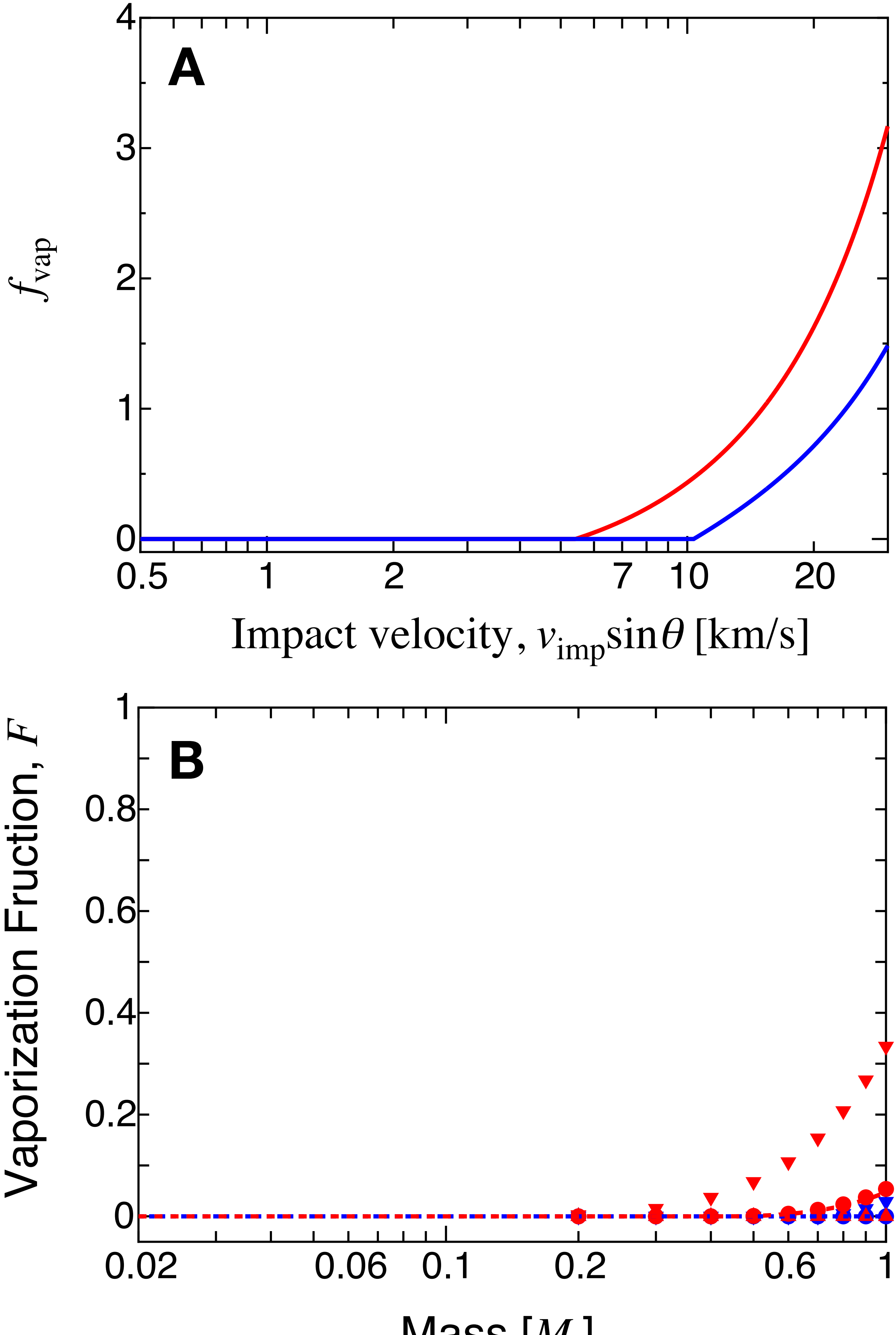

A
$f_{\rm vap}$
0
1
2
3
4
0.5
1
2
7
10
20
Impact velocity, $v_{\rm imp}\sin\theta$ [km/s]
B
Vaporization Fruction, $F$
0
0.2
0.4
0.6
0.8
1
0.02
0.06
0.1
0.2
0.6
1
Mass [$M$]

Fig. 3

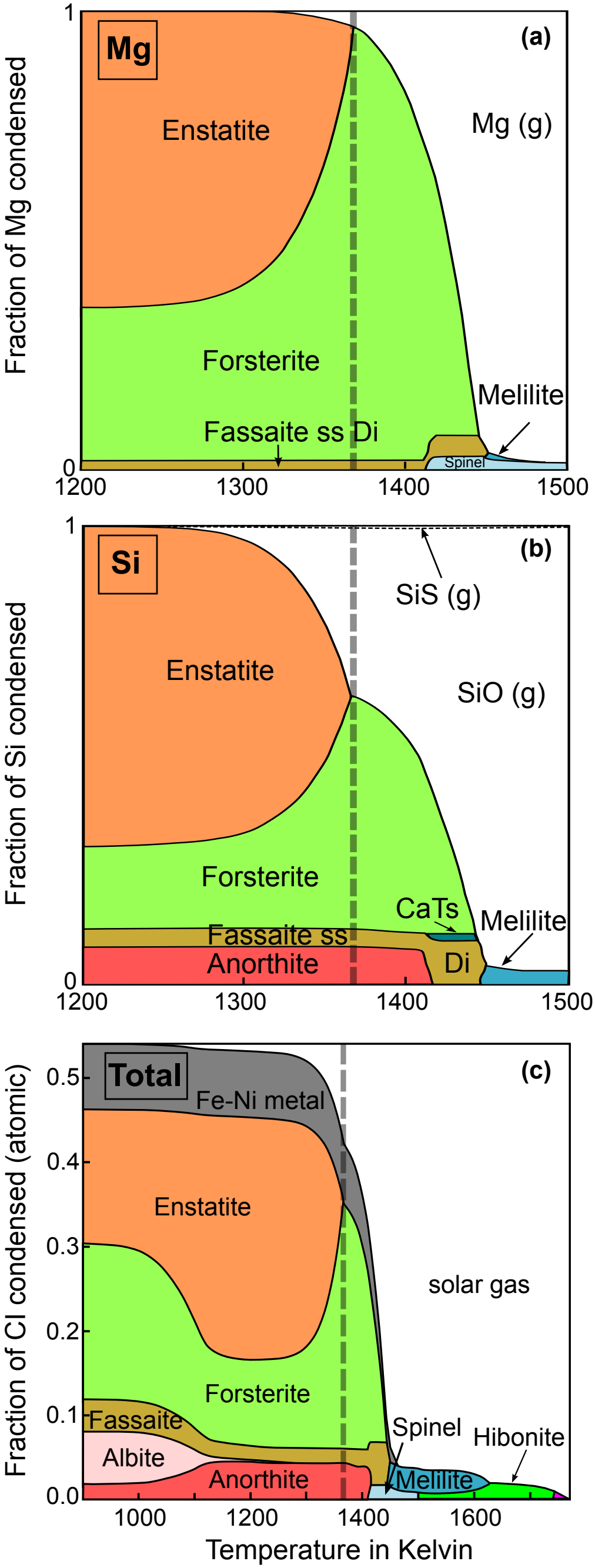

Mg
(a)
Enstatite
Mg (g)
Forsterite
Melilite
Fassaite ss Di
Spinel
Fraction of Mg condensed
1
0
1200
1300
1400
1500
Si
(b)
SiS (g)
Enstatite
SiO (g)
Forsterite
CaTs
Melilite
Fassaite ss
Anorthite
Di
Fraction of Si condensed
1
0
1200
1300
1400
1500
Total
(c)
Fe-Ni metal
Enstatite
solar gas
Forsterite
Fassaite
Spinel
Hibonite
Albite
Anorthite
Melilite
Fraction of CI condensed (atomic)
0.5
0.4
0.3
0.2
0.1
0.0
1000
1200
1400
1600
Temperature in Kelvin

Fig. 4

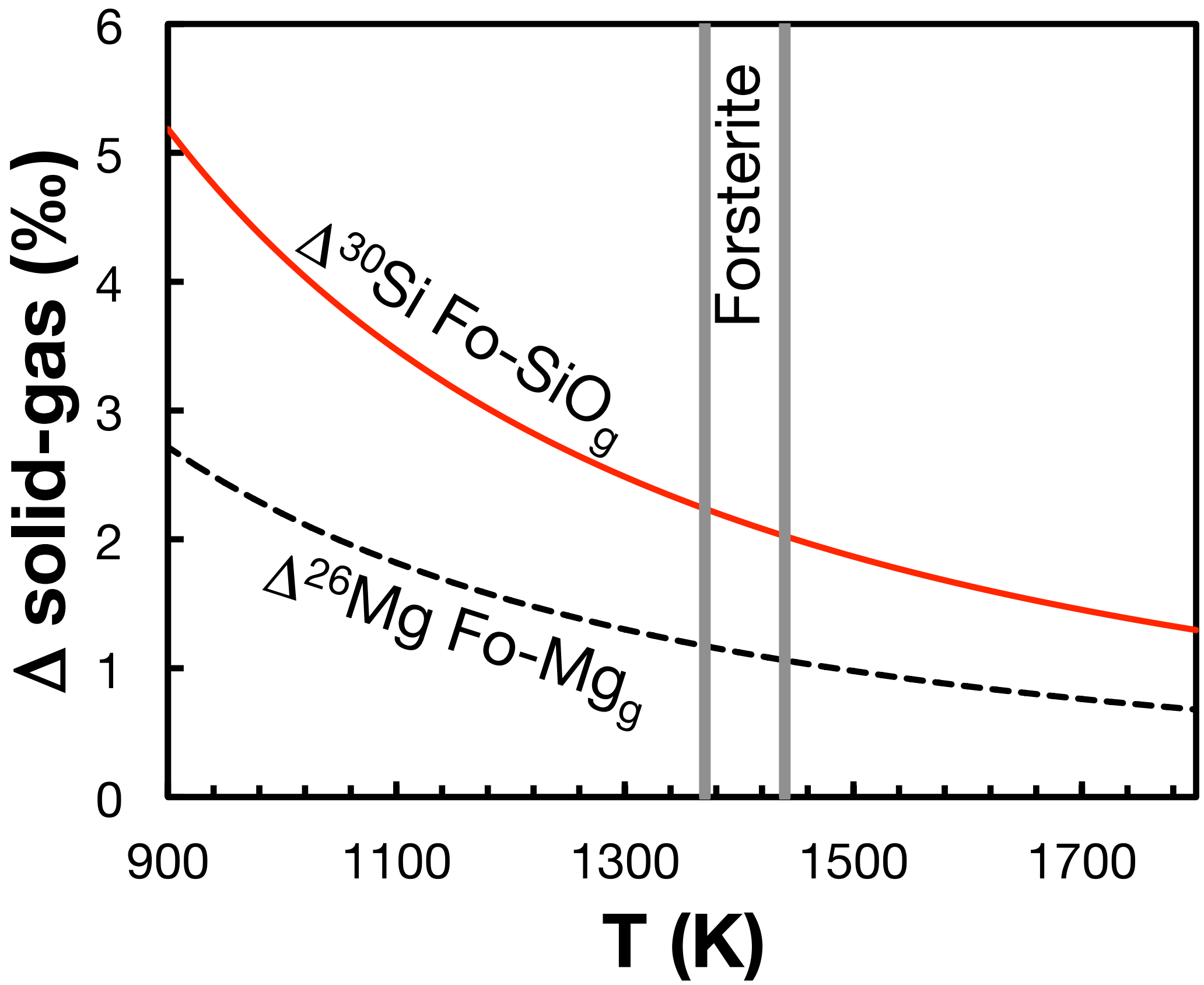
6
5
4
3
2
1
0
Δ solid-gas (‰)
900
1100
1300
1500
1700
T (K)
Forsterite
Δ30Si Fo-SiOg
Δ26Mg Fo-Mgg

Fig. 5

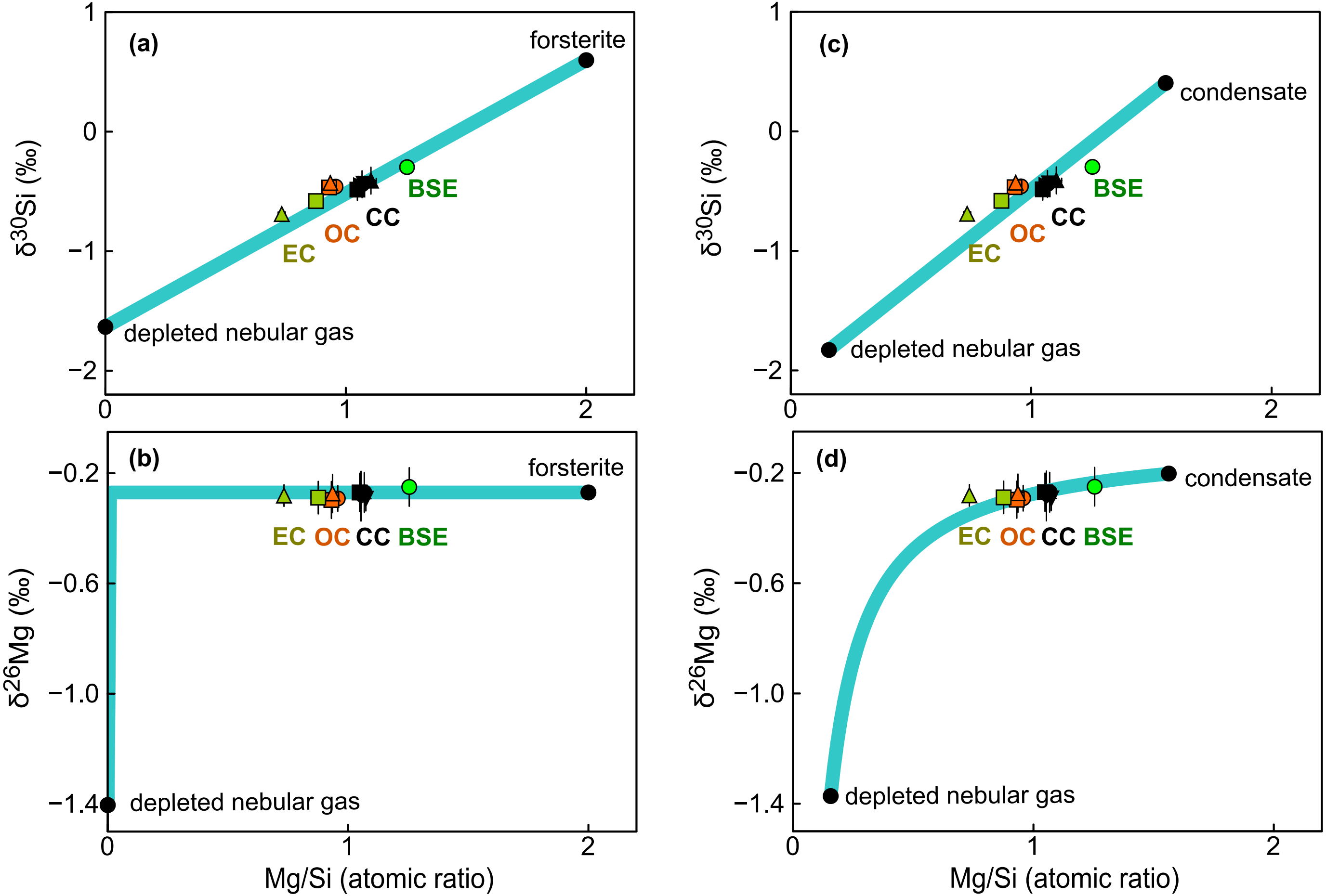
(a)
forsterite
depleted nebular gas
BSE
CC
OC
EC
δ30Si (‰)
(b)
forsterite
depleted nebular gas
EC OC CC BSE
δ26Mg (‰)
Mg/Si (atomic ratio)
(c)
condensate
depleted nebular gas
BSE
CC
OC
EC
δ30Si (‰)
(d)
condensate
depleted nebular gas
EC OC CC BSE
δ26Mg (‰)
Mg/Si (atomic ratio)

Fig. 6

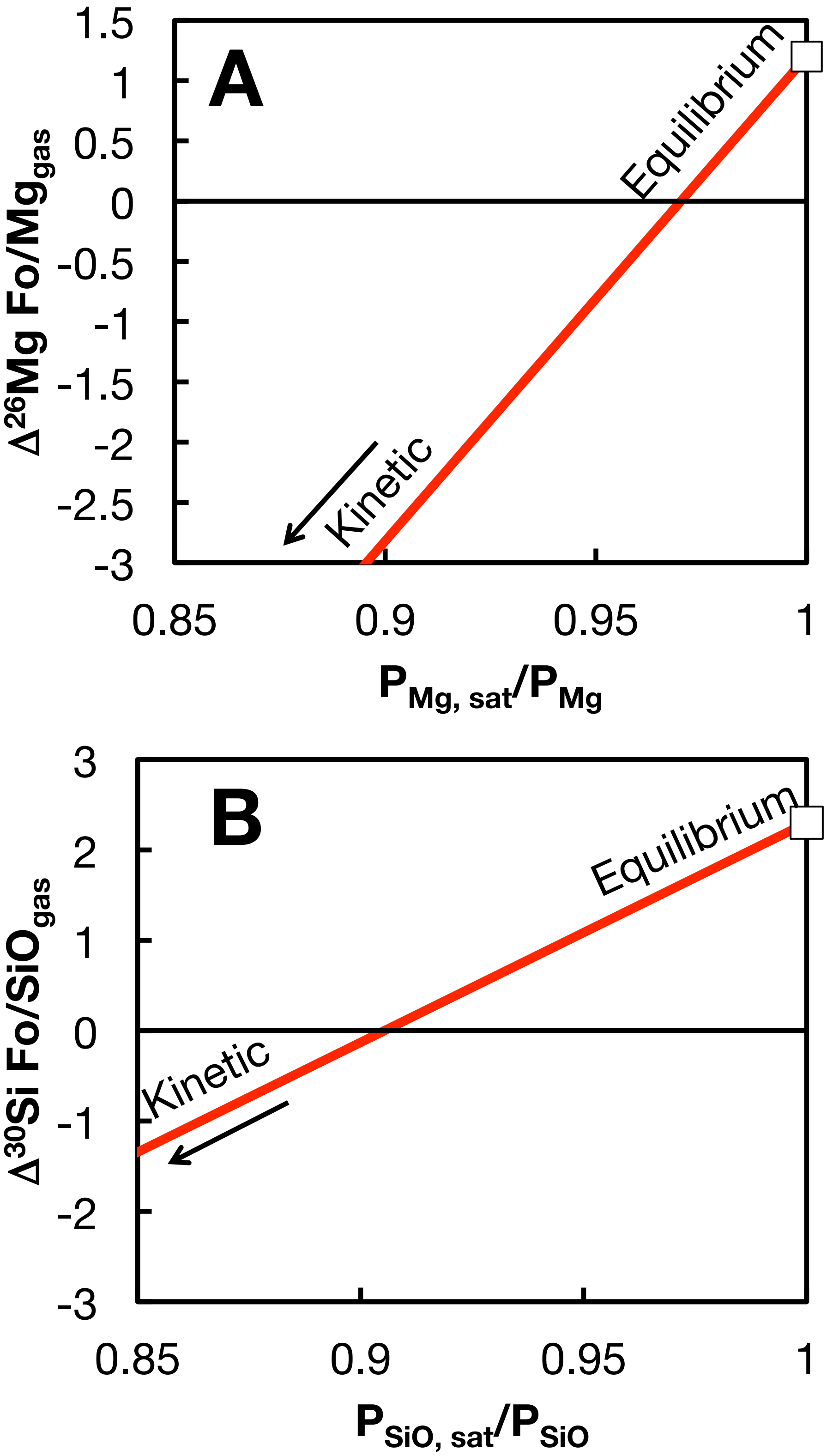
A
Equilibrium
Kinetic
$\Delta^{26}$Mg Fo/Mg$_{gas}$
1.5
1
0.5
0
-0.5
-1
-1.5
-2
-2.5
-3
0.85
0.9
0.95
1
$P_{Mg, sat}/P_{Mg}$
B
Equilibrium
Kinetic
$\Delta^{30}$Si Fo/SiO$_{gas}$
3
2
1
0
-1
-2
-3
0.85
0.9
0.95
1
$P_{SiO, sat}/P_{SiO}$

Fig. 7

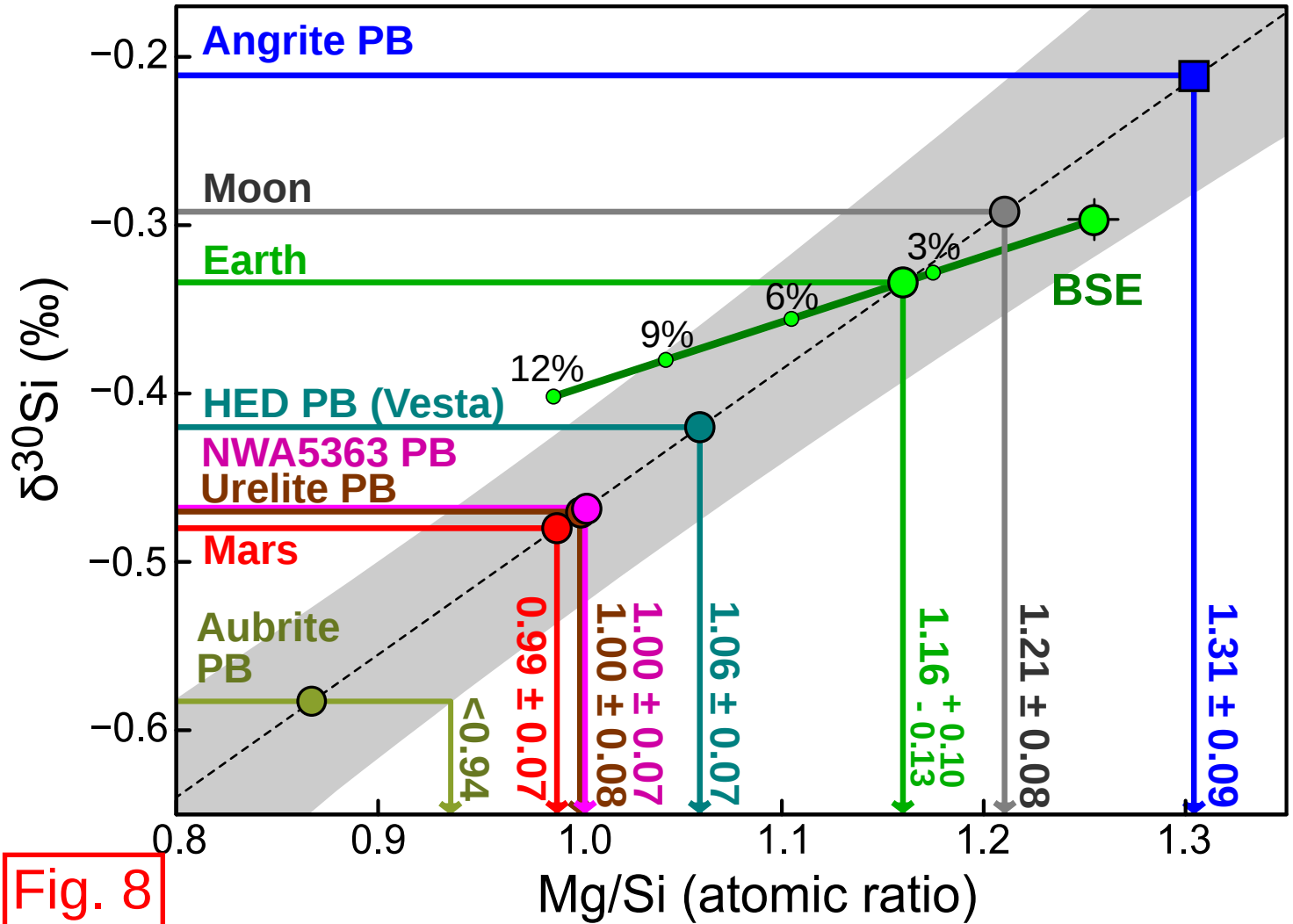

Angrite PB
Moon
Earth
BSE
3%
6%
9%
12%
HED PB (Vesta)
NWA5363 PB
Urelite PB
Mars
Aubrite PB
<0.94
0.99 ± 0.07
1.00 ± 0.08
1.00 ± 0.07
1.06 ± 0.07
1.16 + 0.10 - 0.13
1.21 ± 0.08
1.31 ± 0.09
−0.2
−0.3
−0.4
−0.5
−0.6
δ30Si (‰)
0.8
0.9
1.0
1.1
1.2
1.3
Mg/Si (atomic ratio)

Fig. 8